%% file: Neq2SCFT_VOA.tex
\RequirePackage{pdf14}

\documentclass[11pt]{article}
\usepackage[usenames,dvipsnames,table]{xcolor}
\usepackage[utf8]{inputenc}
\usepackage{graphicx}
\usepackage{caption}
\usepackage{subcaption}
\usepackage{bbold,array,setspace,mathrsfs,amsthm,amsfonts,amsmath,yfonts,dsfont,bbm,colonequals,mathtools}
\usepackage{./aux/jheppub}
\usepackage{afterpage}
\usepackage{xspace}
\usepackage{braket}
\usepackage[normalem]{ulem}
\usepackage[percent]{overpic}
\usepackage{tikz-cd}

\input{./aux/topmatter}

\title{Lectures on chiral algebras of \texorpdfstring{$\mathcal{N} \geqslant 2$}{N >=2} superconformal field theories}

\author{Madalena Lemos}

\affiliation{Theoretical Physics Department, CERN, 1211 Geneva 23, Switzerland.\\
Department of Mathematical Sciences, Durham University, Lower Mountjoy, DH1 3LE Durham, United Kingdom.}

\emailAdd{madalena.lemos@durham.ac.uk}

\preprint{CERN-TH-2020-098}

\bigskip
\abstract{
Any four-dimensional $\mathcal{N} \geqslant 2$ superconformal field theory possess a protected subsector isomorphic to a two-dimensional chiral algebra \cite{Beem:2013sza}. The goal of these lectures is to provide an introduction to the subject, covering the construction of the chiral algebras, the consequences for four-dimensional physics, as well as a brief summary of recent progress.
This is the writeup of the lectures given at the Winter School ``YRISW 2020'' to appear in a special issue of JPhysA.
}


\begin{document}
\setcounter{tocdepth}{2}
\maketitle
\setcounter{page}{1}

\input{sections/1_introduction}
\input{sections/2_2dCFT}
\input{sections/3_chiralalgebra}

\input{sections/4_consequences}
\input{sections/5_outlook}

\input{sections/acknowledgments}
\appendix
\input{sections/A_superalgebra}

\bibliography{./aux/biblio}
\bibliographystyle{./aux/JHEP}

\end{document}

%% file: aux/topmatter.tex
\newtheorem{exercise}{Exercise}
\newtheorem*{exercise*}{Exercise}
\newtheorem{example}{Example}
\newtheorem*{example*}{Example}
\newtheorem{result}{Result}
\newtheorem*{result*}{Result}

\newtheorem*{conj*}{Conjecture}

\def\LT{L^{T}}
\def\Md{{\bar{\MM}}}
\def\ZZ{{\mathcal{Z}}}
\def\mm{{\mathfrak{M}}}

\def\infsum{{\sum\limits_{n=-\infty}^{+ \infty} }}
\def\vac{{|0\rangle}}
\def\state{{|\phi\rangle}}

\def\pd{{\dot{+}}}
\def\md{{\dot{-}}}

\def\hb{{\bar{h}}}

\def\sl{sl}

\def\BB{{\mathcal{B}}}
\def\CC{{\mathcal{C}}}

\def\EE{{\mathcal{E}}}
\def\OO{{\mathcal{O}}}
\def\NN{{\mathcal{N}}}
\def\LL{{l}}

\def\II{{i}}

\def\QQ{Q}
\def\MM{{\mathcal{M}}}
\def\HH{D}
\def\JJ{j}

\def\PP{P}
\def\KK{K}
\def\SS{S}
\def\RR{{R}}

\def\Qt{\tilde{Q}}
\def\St{\tilde{S}}

\def\Lb{\bar{L}}

\def\Mort{\MM_{\perp}}
\def\aa{\alpha}
\def\aad{{\dot{\alpha}}}
\def\bb{\beta}
\def\bbd{{\dot{\beta}}}
\def\gg{\gamma}
\def\ggd{{\dot{\gamma}}}

\def\ddd{\dot{\delta}}

\def\Lb{\bar{L}}
\def\Lh{\hat{L}}

\def\jb{{\bar{j}}}

\def\Qt{\tilde{Q}}
\def\St{\tilde{S}}
\def\ii{i}

\def\sl{sl}

\def\ad{\dot{\alpha}}

\renewcommand{\a}{\alpha}

\newcommand{\D}{\Delta}

\newcommand{\zb}{{\bar{z}}}

\newcommand{\Tr}{\textup{Tr}}

\newcommand{\be}{\begin{equation}}
\newcommand{\ee}{\end{equation}}
\newcommand{\bea}{\begin{eqnarray}}
\newcommand{\eea}{\end{eqnarray}}

\newcommand\qq{\mathbbmtt{Q}}

\newcommand{\beq}{\begin{equation}}
\newcommand{\eeq}{\end{equation}}

\newcommand\eg{\emph{e.g.}}
\newcommand\ie{\emph{i.e.}}

%% file: sections/1_introduction.tex

\section*{Structure of lectures}

These lectures were delivered as five 45 minute lectures in the ``YRISW 2020: A modern primer for superconformal field theories'' school taking place from 9-16 February 2020 at DESY. The exercises throughout these lectures were given in the problem sessions of the school.
The content of these lectures was coordinated with that of the other lecturers \cite{Lorenz,Mario,Abhijit, Bruno}. In particular, these lectures require all of the material covered in \cite{Lorenz}, section~\ref{sec:props} assumes familiarity with the contents of \cite{Abhijit} up to and including section~3. Finally, the examples in sections~\ref{sec:chiralalg} and~\ref{sec:consequences} connect with examples discussed in \cite{Mario,Bruno}.  

\section{Introduction and motivation}

Throughout the lectures of this school \cite{Lorenz,Abhijit,Mario, Bruno} one encounters a large zoo of four-dimensional Superconformal Field Theories (SCFTs). Some of these theories have a conventional Lagrangian description as gauge theories, as several of those appearing in \cite{Mario}, while others  lack a Lagrangian description, such as many of the theories arising as twisted compactifications of the maximally supersymmetric six-dimensional theory in \cite{Bruno}. In fact some of the theories obtained are intrinsically strongly coupled, and have no parameters one can tune to make the theory perturbative in any sense. 
These lectures have shown that the landscape of four-dimensional supersymmetric theories is vast and very rich, however, it is also strongly constrained  by symmetry.
For maximally supersymmetric theories ($\NN=4$) one believes to have a complete classification, \ie, $\NN=4$ Super-Yang-Mills, even though a complete proof is still elusive. One would then hope to solve these theories, the precise meaning of which will be defined shortly. The recently discovered $\NN=3$ theories \cite{Garcia-Etxebarria:2015wns} still have a large amount of supersymmetry, making theory space still very constrained. 
Going down in supersymmetry, the space of $\NN=2$ SCFTs is much richer than $\NN=3$ or $\NN=4$, while remaining constrained enough that one could hope for a  complete classification. This is unlike the case of $\NN=1$ supersymmetry, where at this point a complete classification seems far in the future. These lecture notes will focus precisely on theories with $\NN \geqslant 2$ supersymmetry.

The different lectures in this school present different approaches to classify and study the dynamics of SCFTs in four dimensions. The approach described in these lecture notes also attempts to make progress in  these two directions, by obtaining a subsector of $\NN \geqslant 2$ SCFTs that is captured by a two-dimensional chiral algebra \cite{Beem:2013sza}. These results will apply to any SCFT, irrespectively  of having any type of microscopic description, or any parameter that can be used for perturbation theory. As such, it is useful to describe SCFTs in a purely abstract way, that eschews any such descriptions. These lectures start by introducing such an  abstract operator-algebraic language to describe $\NN \geqslant 2$ SCFTs, the groundwork of which is covered in \cite{Lorenz}. We will then give a quick review of two-dimensional Conformal Field Theories (CFTs) in section~\ref{sec:2dcft} before proceeding to obtain the aforementioned subsector and studying its consequences in sections~\ref{sec:chiralalg} and~\ref{sec:consequences}.

\subsection{Abstract approach to CFTs}

In the lectures \cite{Lorenz,Mario,Bruno} you've encountered various examples of local operators in free and Lagrangian theories, for example 	\emph{Coulomb branch}  and \emph{Higgs branch} operators. In the example of $\NN=2$ superconformal QCD  given in section~1.1 of \cite{Mario} operators of these two types are given as  $\Tr \phi^2$ and the meson operator, $\mm_{\ell_2}^{\ell_1}= (\bar{q}^\dagger)^{\ell_1}_A \bar{q}^A_{\ell_2}$ (see Digression~1.1 of \cite{Mario}), respectively. Given that such a microscopic Lagrangian description is not possible for all theories of interest, with examples of this showing up in all the lectures, we would like to follow an abstract approach, that does not require such a  description. 
From \cite{Lorenz} we have learned how to organize operators in irreducible representations of the (super)conformal algebra of the theory at hand. In specifying a theory we then need to list all superconformal representations that are present (and their respective multiplicities).
The two aforementioned examples then translate into stating that the theory in question -- $\NN=2$ $su(2)$ superconformal QCD --  has a $B_1 \bar{L}[0;0]_{2}^{(0;4)}$ ($\Tr \phi^2$) and a $B_1 \bar{B}_1[0;0]^{(2;0)}_2$ ($\mm_{\ell_2}^{\ell_1}$) superconformal multiplets present.\footnote{We use the same labeling of superconformal multiplets as \cite{Lorenz} which corresponds to the one of \cite{Cordova:2016emh}. In the notation of \cite{Dolan:2002zh} these are $\EE_2$ and $\hat{B}_1$ multiplets respectively.}
Furthermore, we should use all available symmetries to simplify the problem, and thus if the theory has any \emph{global symmetry}, \ie, a symmetry that commutes with all spacetime symmetries (the superconformal algebra),  we will organize operators in representations of such global, or \emph{flavor} symmetry. 

\subsubsection*{Defining a CFT}

We thus take the abstract viewpoint that \emph{a conformal field theory (CFT) is defined by the set of all local operators and their correlation functions}.\footnote{In doing so we willfully ignore non-local operators, such as line or surface defects, that are interesting in their own right, and necessary for a complete description of a quantum field theory \cite{Aharony:2013hda}. However, the subset of local operators is closed on itself when studying correlation functions in flat space, and thus it is a consistent truncation of the \emph{full} CFT spectrum. As such for most of these lectures we will take the definition of a CFT to include only local operators, as a starting point for our analysis, keeping in mind that we will want to extend the set of observables. Including non-local operators is briefly discussed in section~\ref{sec:defects}.}
Our observables are then the correlation functions of local operator. We must first use symmetries to fix their kinematics, to separate the information that is completely fixed by symmetries, from that which is theory dependent, \ie, dynamical.\footnote{For a review on conformal field theories and the conformal bootstrap see the beautiful lectures of \cite{Simmons-Duffin:2016gjk}.}
Starting from one point-functions we see that scale invariance requires
\be 
\langle \OO(x) \rangle \neq 0 \,, \qquad \mathrm{iff} \; \Delta_\OO =0\,,
\label{eq:onepoint}
\ee
and by assumption the unique dimension zero operator in a CFT is the identity operator $\mathbb{1}$.

\begin{exercise}
Taking  $\OO_i$ to be scalar primary operators, show conformal invariance requires
\be 
\langle \OO_i(x_1) \OO_j(x_2) \rangle = \frac{\delta_{\Delta_i - \Delta_j} C_{ij} }{x_{12}^{2\Delta}}\,,
\label{eq:twopoint}
\ee
where $C_{ij}$ is unfixed by symmetry. In a unitary CFT we can pick a basis of operators such that it becomes $C_{ij}=\delta_{ij}$.
Exercise~2 of \cite{Lorenz}  has you show
\be 
\langle \OO_1 (x_1) \OO_2(x_2) \OO_3 (x_3) \rangle = \frac{\lambda_{123}}{|x_{12}|^{\D_1 + \D_2-\D_3}|x_{13}|^{\D_1 + \D_3-\D_2}|x_{23}|^{\D_2 + \D_3-\D_1}}\,,
\label{eq:threepoint}
\ee
where $\lambda_{123}$ is unfixed by symmetry.
\end{exercise}
For spinning operators the number of unfixed constants in the three-point function increases (see \eg, \cite{Simmons-Duffin:2016gjk}), but there is still a finite number of unfixed coefficients for each three-point function.
One-, two- and three-point functions are thus completely fixed by symmetry, up to knowing the operator  operator content of the theory and the three-point couplings $\lambda_{123}$ together with their spinning counterparts. Higher $n$-point functions become non-trivial functions of conformally invariant cross-ratios made from the $n$ positions of operators.

\subsubsection*{The operator product expansion}
CFTs have an \emph{Operator Product Expansion} (OPE) with a finite radius of convergence, which can be used inside correlation functions to compute them,
\be 
\OO_1 (x) \OO_2(x) \sim \sum\limits_{\substack{k \text{ scalar} \\ \text{conformal primaries}}} \frac{\lambda_{12k}}{x^{\Delta_1 + \Delta_2 -\Delta_k}} \left(\OO_k + \beta x^\mu \partial_\mu \OO_k + \ldots \right) + \substack{\text{similar terms for}\\ \text{spinning operators}}\,.
\label{eq:OPE}
\ee
Where we took $\OO_{1,2}$ to be scalar operators and show explicitly only the contrition of an exchanged scalar operator for simplicity. The sums run over conformal primary operators, with the brackets containing the contribution of the primary and all its descendants $\partial_{\mu_1} \ldots \partial_{\mu_n}\OO$. The contribution of all descendant operators are fixed in terms of those of the primary by conformal symmetry.
\begin{exercise}
Compute $\beta$ by using the OPE \eqref{eq:OPE} inside the three-point function \eqref{eq:threepoint}. You will also see why the same $\lambda_{12k}$ was used in the OPE and three-point function. Similarly all subsequent terms would also be fixed, and the same exercise can be repeated for spinning operators.
\label{exercise:OPEkinematics}
\end{exercise}
For this reason the three-point couplings  $\lambda_{ijk}$ are also called \emph{OPE coefficients}.
If the theory is supersymmetric, then further relations may arise between OPE coefficients of different conformal primaries that are related by supersymmetry.

Therefore, to compute any $n-$point function one only needs to know the spectrum of operators and the set of OPE coefficients $\{\lambda_{ijk}\}$ -- this set is often called the \emph{CFT data}.

\subsubsection*{Conformal block decomposition}

Let us go back to four-point functions, and use a double OPE expansion to compute it from the CFT data. Considering four identical scalars for simplicity, conformal symmetry fixes the form of the four-point function as
\be 
\langle \OO(x_1) \OO(x_2) \OO(x_3) \OO(x_4) \rangle = \frac{1}{x_{12}^{\Delta_\OO}x_{34}^{\Delta_\OO}} \, g(u,v)\,,
\qquad \text{where } u=\frac{x_{12}^2 x_{34}^2}{x_{13}^2 x_{24}^3}\,, \;\; v  = \frac{x_{23}^2 x_{14}^2}{x_{13}^2 x_{24}^3}\,,
\label{eq:four-point}
\ee
are conformally invariant cross-ratios, and thus the four-point function can depend non-trivially on them. One can easily check that there are two such cross-ratios by using conformal symmetry to fix the four operators to be on the same plane, and further fix the location of three of the four operators, with the cross-ratios corresponding to the position of the fourth operator.

We can now use a double OPE expansion to decompose the four-point function as
\be 
\langle (\OO(x_1) \OO(x_2)) (\OO(x_3) \OO(x_4)) \rangle = \frac{1}{x_{12}^{\Delta_\OO}x_{34}^{\Delta_\OO}} 
\sum_{\substack{\Delta_k, \ell_k\\ \text{conformal}	\\\text{primaries}}} \lambda_{12k} \lambda_{34k} \; g_{\Delta_k \ell_k}(u,v)\,,
\label{eq:confblockdec}
\ee
where brackets signal the OPE we took, and where we used orthogonality of two-point functions \eqref{eq:twopoint} to write a single sum over exchanged operators $\OO_k$. The sum runs over all conformal primaries $\OO_k$ in the theory, and the functions $g_{\Delta_k \ell_k}(u,v)$, which are  called \emph{conformal blocks}, capture the contributions of all descendants of the primary $\OO_k$. These functions are completely fixed by symmetries, just like the OPE, whose simplest case is the subject of exercise~\ref{exercise:OPEkinematics}.

\subsubsection*{Bootstrapping CFTs}

Consistency of the CFT requires the operator product algebra to be associative, that is if we have a product of three operators, it cannot matter in which order we use the OPE
\be 
\left(\left(\OO_1 (x_1) \OO_2(x_2)\right) \OO_3(x_3) \right) =\left( \OO_1 (x_1) \left(\OO_2(x_2) \OO_3(x_3) \right)\right) \,,
\label{eq:assoc}
\ee
where again brackets are used to denote which OPEs were taken. This requirement, together with imposing unitarity of the CFT, places very stringent constraints on the allowed sets of CFT data. Note that in \eqref{eq:confblockdec} we took one of the possible double OPE decompositions of the four-point function \eqref{eq:four-point}. Associativity of the operator product algebra guarantees that different choices yield the same result.\footnote{One can translate the constraints of associativity in requiring that different OPE decompositions of a given four-point function all give the same result. One then gets a set of functional equations called \emph{crossing equations}. It can be easily shown that associativity of operator product algebra is equivalent to requiring the crossing equations are satisfied  for all possible four-point functions in the theory, see \eg, \cite{Simmons-Duffin:2016gjk}.}

\bigskip

This leads us to the \emph{bootstrap} approach to CFTs \cite{Ferrara:1973yt,Polyakov:1974gs,Mack:1975jr}, namely the hope that symmetries and general consistency requirements, perhaps allied with a few assumptions, are enough to completely \emph{solve} a theory, \ie, obtain its CFT data.  

\bigskip

We see immediately that we have an infinite number of consistency requirements to impose, eq.~\eqref{eq:assoc} for all possible operators in the theory, for infinitely many unknowns -- the CFT data. This is in general a very hard task. It was originally successful in the case of rational $2d$ CFTs, or theories such as Liouville, see \eg, \cite{DiFrancesco:1997nk}. Starting from the work of \cite{Rattazzi:2008pe} this approach has been applied to CFTs in different dimensions and with different (super)symmetries, and has led to a large wealth of results for CFTs, including for strongly coupled theories that have no Lagrangian description, see \eg, \cite{Poland:2018epd} for a recent review. The approach of \cite{Rattazzi:2008pe}  involves constraining, by numerical means, the space of allowed CFT data, often finding that interesting theories lie at the edge of this allowed space, paving the way for their solution.
Different bootstrap inspired approaches to CFTs have appeared in recent years, that frequently involve studying \eqref{eq:assoc} in particular limits that make the problem more tractable.

\subsubsection*{A solvable subsector of $\NN \geqslant 2$ SCFTs}

In these lectures we will discuss one such approach, by finding a protected subsector of any  $\NN\geqslant2$ SCFT that is isomorphic to a two-dimensional chiral algebra \cite{Beem:2013sza}. Two-dimensional chiral algebras are the holomorphic/left-moving part of a two-dimensional CFT -- a meromorphic CFT, which are  more commonly referred to  as Vertex Operator Algebra (VOA) in the mathematical literature.
The main focus of the lectures will be to show the result of \cite{Beem:2013sza}
\begin{result}
$\exists$ map $\chi\,: \; \{4d\; \NN=2\; SCFTs\} \longrightarrow \{ 2d \text{ chiral algebras}\}$
\label{res:map}
\end{result}
Through this map we will be able to use the full power of meromorphic conformal field theories to learn about the physics of four-dimensional SCFTs, albeit only about a subsector of their full operator spectrum. We will be able to compute observables of the SCFT that would be hard to obtain by other methods, especially for non-Lagrangian theories. The goals are two-fold as alluded to before, one would like to have a complete catalog of SCFTs, but also to know more about specific theories, \ie, to completely solve them.

Since other lectures have introduced $\NN=2$ SCFTs in great detail these lectures will start by a short review of two-dimensional chiral algebras in section~\ref{sec:2dcft}. Section~\ref{sec:chiralalg} will show result~\ref{res:map} and obtain the main properties of the map, by seeing what four-dimensional physics implies for the two-dimensional chiral algebra.
In section~\ref{sec:consequences} we will look at the map in the other direction, and see examples of what we can learn about strongly coupled four-dimensional SCFTs from two-dimensional chiral algebras.
Finally in section~\ref{sec:outlook} we list other recent developments in the context of studying chiral algebras of $\NN \geqslant 2 $ SCFTs as well as cases where the map~\ref{res:map} exists in different dimensions.

\subsubsection*{A note on conventions}
The conventions used in these lectures are chosen to match those of \cite{Lorenz,Abhijit,Mario,Bruno} and differ from those of \cite{Beem:2013sza} at times.
The conventions for labeling superconformal multiplets are those of \cite{Cordova:2016emh}, and following this reference, representations of $su(2)_R$ and Lorentz spins are given by specifying Dynkin labels, while the $u(1)_r$ charge is $r_{here}=r_{\text{\cite{Cordova:2016emh}}}= -2 r_{\text{\cite{Dolan:2002zh}}}$.

%% file: sections/2_2dCFT.tex

\section{Two-dimensional conformal field theories and chiral algebras}
\label{sec:2dcft}

In this section we briefly review some aspects of two-dimensional CFTs and chiral algebras needed for the following sections. There are numerous very nice reviews on two-dimensional CFTs and chiral algebras, \eg, \cite{Ribault:2014hia,DiFrancesco:1997nk,Ginsparg:1988ui}. In two dimensions the conformal algebra is enhanced to an infinite dimensional algebra, as discussed in \cite{Lorenz}, and in this section we will see the consequences of that enhancement.

\subsubsection*{Global conformal algebra}
The global algebra in two-dimensions is the same that exists in higher dimensions, it is generated by 
\be 
P_\mu\,, \HH\,, M_{\mu \nu}\,, K_\mu\,,
\ee
and has the commutation relations summarized in appendix \ref{app:confalgebra}. Since we are in two dimensions let us change to complex coordinates
\be 
z= x^1 - \ii x^2\,, \qquad \zb = x^1 + \ii x^2\,, \qquad \partial_z = \frac{1}{2} \left(\partial_1 + \ii \partial_2 \right)\,, \qquad \partial_{\zb} = \frac{1}{2} \left(\partial_1 - \ii \partial_2 \right)\,,
\ee
where in Euclidean signature one must take $z^* = \zb$, and in Lorentzian one takes $z$ and $\zb$ independent real variables. In what follows it is convenient to think of $z$ and $\zb$ as independent complex coordinates, keeping in mind that the relation $z^* = \zb$ picks the Euclidean physical slice.

We can now re-write the global algebra as 
\be
\label{eq:globalalg}
\begin{aligned}
L_{+1}&=\frac{1}{2}\left( K^{1}-\ii K^{2}\right) \,, \qquad &\Lb_{+1}&=\frac{1}{2}\left( K^{1}+\ii K^{2}\right) \,,\\
L_{0}&=\frac{1}{2}\left( D - \ii M_{12}\right) \,,  &\Lb_{0}&=\frac{1}{2}\left( D+\ii M_{12}\right) \,,\qquad \\
 L_{-1}&=\frac{1}{2}\left( P_{1}+\ii P_{2}\right)\,,   &\Lb_{-1}&=\frac{1}{2}\left( P_{1}-\ii P_{2}\right) \,,
\end{aligned}
\ee
such that the left and right moving algebras factorize, with the only non-vanishing commutation relations reading
\be 
[L_{m},L_{n}]=(m-n) L_{m+n}\,, \; \text{for } m,n=0,\pm1\,, \qquad [\Lb_{m},\Lb_{n}]=(m-n) \Lb_{m+n}\,, \; \text{for } m,n=0,\pm1\,.
\ee
With this re-writing we see explicitly the $sl(2) \times \overline{sl(2)}$ global conformal algebra, \ie, the generators that are defined globally on the Riemann sphere (the complex plane with the point at infinity added).

\subsubsection*{Infinite dimensional conformal symmetry}
As shown in \cite{Lorenz} the local symmetry algebra is infinite dimensional, corresponding to analytic coordinate transformations
\be 
\label{eq:analcoordtransf}
z \to f(z)\,, \qquad \zb \to \bar{f}(\zb)\,,
\ee
with infinitesimal transformations given by $z \to z + \epsilon(z)$ and similarly for $\zb$, where analyticity allows us to write 
\be 
\epsilon(z) =\infsum  z^{n+1} \epsilon_n\,.
\label{eq:epsz}
\ee

In \cite{Lorenz} the geometrical action of the conformal algebra was written down, now let us get a current generating these transformations. We consider local theories, which implies there exists  a stress tensor, \ie, a conserved spin two symmetric traceless operator $T_{\mu \nu}$.

\subsubsection*{The stress tensor and Virasoro symmetry}

\begin{exercise}
Write $T_{\mu \nu}$ in complex coordinates $z\,, \zb$ and show that traceless imposes $T_{z \zb}= T_{\zb z}=0$, while conservation imposes $\partial_{\zb} T_{zz}=0$ and $\partial_{z} T_{\zb\zb}=0$.
\end{exercise}

Therefore we can define the holomorphic and anti-holomorphic stress tensors respectively as
\be 
T(z) = -2 \pi T_{zz}\,, \qquad \bar{T}(\zb) = -2 \pi  T_{\zb\zb}\,,
\ee
where the factors of $-2 \pi$ are conventional.

We can now define a current and charge for the conformal transformations
\be 
Q_\epsilon = \frac{1}{2 \pi \ii} \oint dz \epsilon(z) T(z)\,,
\ee
where $\epsilon(z)$ is the infinitesimal coordinate transformation.
The Laurent expansion of the stress tensor allows us to define the modes $L_n$ by 
\be 
T(z) = \sum\limits_{n=-\infty}^{+ \infty} z^{-n-2} L_n\,, \qquad L_n= \frac{1}{2 \pi \ii} \oint dz z^{n+1} T(z)\,,
\label{eq:Tmodeexp}
\ee
and similarly for the anti-holomorphic stress tensor $\bar{T}(\zb)$, with modes $\Lb_n$.
The charge now reads
\be 
Q_\epsilon = \infsum \epsilon_n L_n\,, \qquad Q_{\bar{\epsilon}} = \infsum \bar{\epsilon}_n \Lb_n\,,
\ee
and we see that $L_n$ and $\Lb_n$  are the generators of local conformal transformations on the Hilbert space.  The modes of the stress tensor obey the \emph{Virasoro} algebra
\be 
\begin{split}
 [L_m,L_n] &= (m-n) L_{m+n} + \frac{c}{12}(m^3  -m) \delta_{m+n,0}\,, \\
  [\Lb_m,\Lb_n] &= (m-n) \Lb_{m+n} + \frac{c}{12}(m^3  -m) \delta_{m+n,0}\,,\qquad  [L_n,\Lb_m]=0\,,
\end{split}
 \label{eq:Viralgebra}
\ee
where for $m,n=0,\pm1$ we recover the global algebra.
This algebra has a central extension, the \emph{central charge} $c$, which is a number that commutes with all $L_n$ and $\Lb_n$.
This algebra corresponds to the unique central extension of the Witt algebra, \ie, the algebra of conformal transformations acting on functions. The central extension appears from the fact that a symmetry algebra only needs to have a projective action on states of the Hilbert space, and the projective action of a symmetry is the same as the action of the centrally extended symmetry, see \eg, \cite{Weinberg:1995mt}. We can immediately see the need for a non-zero central extension by computing the OPE of two stress-tensors from the above commutation relation
\be 
T(z) T(0) \sim \frac{c/2}{z^4} + \frac{2 T(0)}{z^2} + \frac{\partial T(0)}{z}+ \text{regular terms}\,,
\label{eq:STOPE}
\ee
and similarly for $\bar{T}(\zb)$.
We see that $c$ appears as the two-point function of the stress tensor, and thus cannot be zero in a unitary theory if $T(z) \neq 0 $.

\begin{exercise}
Obtain \eqref{eq:Viralgebra} from the OPE \eqref{eq:STOPE} by evaluating
\be 
 [L_n,L_m] = \frac{1}{(2 \pi \ii)^2} \oint_0 dw w^{m+1} \oint_w dz z^{n+1} T(z)T(w)\,,
 \ee
 Start by showing that the commutator equals the expression above by computing the commutator of the two charges, $L_m$, $L_n$, and remembering that we are working in radial ordering as explained in \cite{Lorenz}.
\end{exercise}
 
\subsubsection*{Hilbert Space, $sl(2)$ and Virasoro primaries }
 
We now want to look at the Hilbert space of states. In these lectures we will assume there exists a unique vacuum state $\vac$, which is invariant under global conformal transformations,\footnote{In \cite{Bruno} this assumption will be lifted for Liouville.} and thus
 \be 
 L_{\pm1} \vac\,, \qquad L_0 \vac\,, \qquad \Lb_{\pm1} \vac\,, \qquad \Lb_0 \vac\,,
 \ee
 Also, recall if we want the state $\lim\limits_{z,\zb \to 0} T(z,\zb) \vac$ to be well defined we see from \eqref{eq:Tmodeexp} that we must have also
 \be 
 L_{n \geqslant -1} \vac = 0 \,, \qquad \Lb_{n \geqslant - 1} \vac =0\,.
 \ee
 
 We now want to organize operators in representations of the conformal algebra, as conformal primaries and descendants.  In  \cite{Lorenz} the highest weights of the representations of the global conformal algebra obtained were called conformal primaries. Since in two dimensions the conformal algebra is larger, it is helpful to make a distinction between primaries of the \emph{global conformal algebra}, which are called \emph{quasi-primaries} or \emph{$sl(2)$-primaries} in the $2d$ CFT literature, and \emph{Virasoro} primaries, which are primaries under the infinite dimensional algebra. Let us start by reviewing the former.
 
Using  \eqref{eq:globalalg} we see that an $sl(2)$-primary as defined in \cite{Lorenz}
 ($K_\mu \state =0$, $D \state = \Delta_\phi \state$, with $\state$ transforming in some representation of spin $\ell$ of the Lorentz group) satisfies
 \be 
 \label{eq:sl2prim}
 L_{+1}\state = 0\,, \qquad L_0 \state = h_\phi \state\,, \qquad \Lb_{+1} \state =0\,, \qquad \Lb_0 \state = \bar{h}_\phi \state\,.
 \ee
We have the following relation between the holomorphic and anti-holomorphic dimensions ($h$ and $\bar{h}$ respectively), and the conformal dimension, $\Delta$, and spin, $\ell$,
\be 
\Delta_\phi = h_\phi + \bar{h}_\phi\,, \qquad \ell = h_\phi - \bar{h}_\phi\,.
\ee 
Eq.~\eqref{eq:sl2prim}, combined with \eqref{eq:Tmodeexp}, implies that the stress tensor has the following OPE with an $sl(2)$-primary
 \be 
 T(z) \phi(0) \sim \underbrace{\ldots}_{\substack{L_{n>1}\state \text{ is not}\\ \text{required to vanish} }} + \frac{\overbrace{0}^{L_{+1}\state =0}}{z^3} + \frac{h_\phi \phi(0)}{z^2} + \frac{\partial\phi(0) }{z} + \text{regular}\,,
 \ee 
where we used the fact that the stress tensor generates the geometrical conformal algebra, \ie, $[L_{-1},\phi(0)]= \partial \phi(0)$. Note that from \eqref{eq:STOPE} we recognize the (anti-) holomorphic stress tensor as an $sl(2)$ primary operator with  dimensions $h=2$, $\bar{h}=0$ ($h=0$, $\bar{h}=2$).
Descendant operators are obtained acting with $L_{-1}$ and $\Lb_{-1}$ an arbitrary number of times, thus constructing the full module.

Now we can organize operators into bigger representations of the full Virasoro algebra.  Note that $[L_0,L_n]=-n L_n$, so generators $L_n$ with $n>0$ lower the conformal dimension of operators. We want the conformal dimensions to be bounded from below, so in a given module there will be a \emph{Virasoro primary} state satisfying 
\be 
L_{n >0} \state = 0\,, \qquad L_0 \state = h_\phi \state\,, \qquad \Lb_{n >0} \state = 0\,, \qquad \Lb_0 \state =\bar{h}_\phi \state\,,
\ee
whose OPE with the stress tensor reads
 \be 
 T(z) \phi(0) \sim 0 + \frac{h_\phi \phi(0)}{z^2} + \frac{\partial\phi(0) }{z} + \text{regular}\,.
 \ee 
Virasoro descendants are obtained by acting with arbitrarily many $L_{-n}$, $\Lb_{-n}$ with $n\geqslant1$. A state obtained by $L_{-n_1} \ldots L_{-n_n}\state$ will have holomorphic dimensions $h_\phi + \sum n_i$, and similarly for the action of $\Lb_n$.
These modules group together infinitely many $sl(2)$ primaries, allowing for $2d$ CFTs with finitely many Virasoro primaries, although an infinite number of $sl(2)$ primaries, such as the minimal models.
It is important to highlight that for special values of $h_\phi$ and $c$ null states can appear in the module considered above, and to obtain an irreducible representation one must quotient by the null state, similarly to what is described in \cite{Lorenz}.

\medskip 

In these lectures we will see both $sl(2)$ and Virasoro primaries making an appearance, so it is important to keep the distinction in mind.
When using the double OPE expansion to decompose a four-point function in conformal blocks one can now use a bigger symmetry algebra, organizing the operators in representations of Virasoro, and writing \emph{Virasoro conformal blocks}. These blocks encode the contribution of all Virasoro descendants of a given Virasoro primary, similarly to the conformal blocks of \eqref{eq:confblockdec} that in $2d$ encode the $sl(2)$ descendants of a given $sl(2)$ primary.

\medskip

Even though the Virasoro algebra, and the stress tensor, factorized in a holomorphic and anti-holomorphic piece which allowed us to discuss both rather independently, generic operators of $2d$ conformal field theories will feature both an holomorphic and anti-holomorphic dependence. The spectrum of the CFT will be given by
\be 
S=\bigoplus\limits_{\mathcal{R},\mathcal{R}'}  m_{\mathcal{R},\mathcal{R}'} \mathcal{R} \otimes \bar{\mathcal{R'}}\,,
\ee
where $m_{\mathcal{R},\mathcal{R}'}$ are multiplicities and $\mathcal{R}$ ($\bar{\mathcal{R'}}$) is a representation of the holomorphic (anti-holomorphic) Virasoro algebra.
A generic operator will have a mode expansion given by
\be 
\phi(z,\zb) = \sum\limits_{m,n=-\infty}^{+\infty} \phi_{m,n} z^{-m-h} \zb^{-n-\bar{h}}\,,
\ee
with only conserved currents, which obey $\partial_\zb \phi(z,\zb)=0$, having no $\zb$ dependence.
The CFTs encountered in \cite{Bruno} will be of this type, correlation functions only make sense once you have both the holomorphic and anti-holomorphic parts, \ie, they depend on both $z$ and $\zb$, and one needs to recall that in Euclidean signature one must set $z^*=\zb$.

\begin{exercise}{Affine Kac Moody current algebras\\}
\label{ex:AKM}
Consider a conserved current $J_\mu^A (z,\zb)$, which transforms in the adjoint of a flavor symmetry algebra, with $A$ an adjoint index. Show that conservation implies the current factorizes holomorphically giving a Virasoro primary operator with $h=1$, $\hb=0$ and one with $\hb=1$, $h=0$. Let us focus on the holomorphic part, the OPE between two currents is given by
\be 
J^A(z) J^B(0) \sim \frac{k_{2d}}{z^2} + \frac{\ii f^{AB}_{\phantom{AB}C} J^C(0)}{z} + \text{regular terms}\,,
\ee
where $f^{AB}_{\phantom{AB}C}$ are the structure constants of the flavor algebra.
Writing the mode expansion as
\be 
J(z) = \sum\limits_{n=-\infty}^{+\infty} z^{-n-1}J_n\,,
\ee
show that this implies the following commutation relation
\be 
\left[J_m^A,J_n^B\right] = \ii f^{AB}_{\phantom{AB}C} J^C_{m+n} + k_{2d} \, m\, \delta^{AB} \delta_{m+n,0}\,,
\ee
whose zero mode algebra we recognize as the flavor Lie algebra.
\end{exercise}

\subsection{Chiral algebras}

What will make an appearance in these lectures are instead \emph{chiral} algebras, or holomorphic conformal field theories, where all operators are conserved currents. This means that in the discussions above we only need to consider the holomorphic part. Correlation functions will now be meromorphic functions of $z$, and will be single valued on their own, without needing to add any $\zb$ dependence. This constrains the dimensions of the operators in the theory -- we will see that coming from four dimensions these conditions are automatically true.

\subsubsection*{OPE and normal ordering}
The two final ingredients we need to elaborate on are the OPE and normal ordering.
Thanks to meromorphicity we can compute any correlation function simply by knowing its singularities, which are controlled by taking the various OPE limits inside the correlation function. Furthermore, in the OPE between two operators
\be 
\OO_1(z_1) \OO_2(0) \sim \sum_i \frac{\lambda_{12i} \OO_i(0)}{z^{h_1+h_2-h_i}}\,,
\ee
we only need to know the singular terms to be able to fix correlation functions. Note that unitarity requires $h_i \geqslant 0 $, thus bounding the strength of the singularity.\footnote{The chiral algebras we will consider are not unitary, however we will see that all states have positive $h$, unless one inserts defects in the four-dimensional SCFT.}

The \emph{normal-ordered} product of two operators is defined as the first regular term in the OPE, namely as
\be 
\left(\OO_1 \OO_2 \right)(0) \colonequals \lim\limits_{z\to 0} \left(\OO_1(z) \OO_2(0) - \text{singular terms}\right)\,.
\ee

\begin{example}
The OPE between the stress tensor and an $sl(2)$-primary is given by
\be 
T(z) \OO(0) = \underbrace{\ldots + \frac{h_\OO \OO(0)}{z^2}+ \frac{\partial \OO(0)}{z}}_{\text{singular piece}} + \left(T\OO\right)(0) + \sum\limits_{k=1}^{+\infty} \frac{z^k}{k!} \left(\partial^k T\OO\right)(0)\,,
\ee
where the summand was obtained by taking the Taylor expansion of $T(z)$ around $\OO$, and which produces Virasoro descendants of $\OO$.
\end{example}

OPEs between normal-ordered products of operators can then be obtained from the OPEs of the operators making up the normal-ordered product, by generalized Wick contractions, see \eg, chapter~6 of \cite{DiFrancesco:1997nk}. As such it suffices to know the singular OPEs between \emph{strong generators} of the chiral algebra, \ie, operators that cannot be written as normal-ordered products (with or without derivatives) of other operators. The chiral algebras we will encounter appear to all be strongly finitely generated, and so one only needs to specify a finite set of generators and their singular OPEs to know any $n$-point function. This will allow us to compute a subsector of correlation functions of four-dimensional $\NN\geqslant 2$ SCFTs from knowledge of the strong generators of the chiral algebra, and their singular OPEs. It is often the case that, for a given set of strong generators, associativity of the operator product algebra \eqref{eq:assoc} is powerful enough to completely fix the singular OPEs, perhaps up to a few coefficients, such as the central charge.

%% file: sections/3_chiralalgebra.tex
\section{Chiral algebra of \texorpdfstring{$4d$ $\NN \geqslant 2$}{4d N>=2} SCFTs}
\label{sec:chiralalg}

\subsection{The claim}

Before going into technical details, let us state the claim of \cite{Beem:2013sza} that we will be proving in this section:

\begin{result}
$\exists$ a subsector of local operators in any four-dimensional $\NN \geqslant 2$ SCFT that is isomorphic to a two-dimensional chiral algebra, or vertex operator algebra.
\label{res:subsector}
\end{result}

\emph{What happens?}

\begin{itemize}
\item Pick a plane $\mathbb{R}^2 \subset \mathbb{R}^4$: Let us pick the $x_3-x_4$ plane and give it coordinates $z$, $\zb$ according to
\end{itemize}
\be 
z= x_3 - \ii x_4 = - x^{+ \pd}\,, \qquad \zb = x_3 + \ii x_4 = x^{- \md}\,.
\ee
\begin{itemize}
\item Now we restrict the local operators, $\OO_1(z_1,\zb_1)\,,\; \ldots\,,\; \OO_n(z_n,\zb_n)$ to lie on this plane, and
\item take ``special'' operators -- operators in the subsector alluded to in the result~\ref{res:subsector} -- that belong in certain short multiplets of the superconformal algebra. These operators transform in non-trivial irreducible representations of $su(2)_R$.
\end{itemize}
Let us use  $I_i$ to collectively denote the $su(2)_R$ indices of the operator $\OO_i(z_i,\zb_i)$.
The claim is that if we contract these $su(2)_R$ indices with a specific (known) function  $u_{I_i}(\zb_i)$,\footnote{For an $su(2)_R$ doublet $u_i(\zb)= (1, - \zb)$.} then
\be 
u_{I_1}(\zb_1) \ldots u_{I_n}(\zb_n) \langle \OO_1^{I_1}(z_1,\zb_1)\; \ldots\; \OO_n^{I_n}(z_n,\zb_n) \rangle = f(\{z_i\})\,,
\ee
is a correlator of a two-dimensional chiral algebra.\footnote{Hints of the existence of this subsector were already present in \cite{Dolan:2001tt,Nirschl:2004pa,Dolan:2004mu}, where the authors used supersymmetric Ward identities to fix the four-point function of four half-BPS $B_1 \bar{B}_1[0,0]_R^{(R;0)}$ operators. The four-point function admits a decomposition in the various $R-$ symmetry representations that can appear in the tensor product of the representations of the external operators. In each of these $R-$symmetry channels one has a function of the two conformally-invariant cross-ratios of four points. Supersymmetric Ward identities, however, impose a set of relations among these functions, and in solving said Ward identities a function of a single variable appears. The function is exactly $f(\{z_i\})$ and one can understand its appearance as performing the construction just described. The authors also found that the crossing equations for these four-point functions had a decoupled equation that only involved the single variable function -- corresponding to the crossing equation in chiral algebra.}
To specify $f(\{z_i\})$ we only need to know its singularities, which arise  from those of the four-dimensional correlator, corresponding to the different OPE limits of that correlator.  We thus get to fix the \emph{full} $f(\{z_i\})$ from knowing only the singular terms of these $4d$ OPE limits  (after performing the twist by $u_{I}(\zb)$), instead of needing the full infinite set of operators exchanged in the $4d$ OPE. Furthermore, after fixing  $f(\{z_i\})$ we can recover an \emph{infinity} of $4d$ OPE coefficients, corresponding to all operators in the protected subsector appearing in each of these OPEs, including the non-singular contributions.\footnote{A subtlety that will be explained in section~\ref{sec:props} is that the identification of the four-dimensional operator that corresponds to a given $2d$ one  can be ambiguous.}

We are thus obtaining the map \ref{res:map} between four-dimensional SCFTs and two-dimensional chiral algebras. This map has interesting consequences that we will explore throughout the lectures: $2d$ chiral algebras have very rigid structures, so they can be used to obtain new results about strongly coupled $4d$ SCFTs. Moreover, some physically motivated results, such as dualities, \ie, different presentations of the same four-dimensional physics,  are not obvious from the chiral algebra point of view, and so implications go both ways.

Now that the claim is introduced let us explain how/why it is true following \cite{Beem:2013sza}, and then we will study the properties of this map.

\subsection{Cohomological construction}

On the $x_3-x_4$ plane, that we call the \emph{chiral algebra plane}, we have the action of an $\sl(2) \times \overline{sl(2)}$ algebra, generated by\footnote{Note that in section~\ref{sec:2dcft} we had a two-dimensional CFT with directions $x_{1,2}$, but now since we are  defining a two-dimensional plane inside a four-dimensional CFT we will write the two-dimensional algebra in the $x_{3,4}$ directions.}
\be
\begin{alignedat}{4}
&2 L_{-1}= \PP_{+ \pd} \,, \qquad  2 L_{+1}& =  \KK^{\pd +} \,, \qquad 2 L_0&= \HH + \MM\,, \\
&2 \Lb_{-1}= -\PP_{- \md} \,, \qquad 2 \Lb_{+1} &= - \KK^{\md -} \,, \qquad 2 \Lb_0&= \HH - \MM\,,
\label{eq:2dsl2}
\end{alignedat}
\ee
where 
\be 
P_{+ \dot{+}} =P_{3} + \ii P_4\,,
\qquad
P_{- \dot{-}} = -\left(P_{3} - \ii P_4\right)\,, \qquad
K^{\dot{+} +} =K_{3} - \ii K_4\,,
\qquad
K^{ \dot{-} -} = -\left(K_{3} + \ii K_4\right)\,,
\ee
and $\MM$ are rotations on the chiral algebra plane
\be 
\MM  = \MM_{+}^{\phantom{+}+} - \Md_{\pd}^{\phantom{\pd}\pd}= \MM_{+}^{\phantom{+}+} + \Md^{\md}_{\phantom{\md}\md} \,.
\ee
The relation to the $4d$ generators was chosen such that the $L_{n}$ and $\Lb_{n}$  obey the standard  $\sl(2) \times \overline{sl(2)}$ algebra
\be 
[L_{m},L_{n}]=(m-n) L_{m+n}\,, \; \text{for } m,n=0,\pm1\,, \qquad [\Lb_{m},\Lb_{n}]=(m-n) \Lb_{m+n}\,, \; \text{for } m,n=0,\pm1\,.
\ee
Some of the fermionic generators also preserve the chiral algebra plane, namely  $\QQ_{-}^\II$, $\Qt_{\II \md}$, $\SS^-_i$ and $\St^{\II \md}$. Altogether this forms an $sl(2)\oplus sl(2|2)$ subalgebra of $su(2,2|2)$ on the chiral algebra plane. There is a central element, \ie, an element that commutes with the full superconformal algebra on the plane\footnote{Note that $r_{here}=-2r_{\text{\cite{Beem:2013sza}}}$.}
\be 
\ZZ = -\frac{r}{2} + \Mort \,, \qquad\text{where} \quad \Mort= \MM_+^{\phantom{+}+}+\Md_{\pd}^{\phantom{\pd}\pd}\,,
\ee
$r$ is the $u(1)_r$ generator, and $\Mort$ are rotations on the plane orthogonal to the chiral algebra plane.

\begin{exercise}
The conventions for the superconformal algebra in $4d$ are given in appendix \ref{app:SCA}, using \eqref{eq:2dsl2} obtain the $sl(2)\oplus sl(2|2)$  subalgebra on the chiral algebra plane.
\end{exercise}

This is just the \emph{global} part of the $2d$ conformal algebra,  and there is no Virasoro enhancement yet -- we are just considering a subalgebra of the four-dimensional algebra.

\subsubsection*{Meromorphicity}

We now want to find something meromorphic, and we will do so by passing to the cohomology  of a suitably chosen supercharge. We would like to find a supercharge $\qq$ such that
\begin{enumerate}
\item $\qq^2=0$, and we will want to obtain its \emph{cohomology}, \ie, operators, $\OO$, that are $\qq-$\emph{closed} ($\qq \OO=0$) but not $\qq-$\emph{exact} ($\OO \neq \qq X$) -- this will be our subsector of operators.
\item We want to keep a $z$ dependence, so we want to be able to translate operators in the $z$ direction and have them remain in cohomology. In fact we want to preserve the \emph{full} $sl(2)$, so we require $[\qq, L_n]=0$, $n=\pm1,0$.
\item We want the anti-holomorphic dependence, \ie, $\overline{sl(2)}$ to drop out, so it should be $\qq-$exact such that it is trivial in cohomology. We that want $\Lb_n =\{\qq, \text{something}\}$, such that $\Lb_{-1} = \partial_\zb$ is $\qq-$exact.
\end{enumerate}

\begin{exercise}
Show that if  $\Lb_{-1} = \partial_\zb$ is $\qq-$exact then $\partial_\zb \langle \OO_1 \cdots \OO_n \rangle =0$, where the operators $\OO_i $ are in cohomology.
\end{exercise}

Note that properties~1 and~2 are automatically true, since supercharges are nilpotent and the supersymmetry is only in the anti-holomorphic sector, thus it commutes with the holomorphic conformal generators.\footnote{This last fact explains why we picked an $sl(2)\oplus\sl(2|1)$ subalgebra of $su(2,2|2)$ instead of other choice for a maximal subalgebra of $su(2,2|2)$ that preserves the plane which would be an $sl(2|1)\oplus \sl(2|1)$.}
Property~3 will not be satisfied by any choice, so we will slightly modify it by requiring only that there exists an $\widehat{sl}(2)$ obtained from $\overline{sl(2)}$ that is $\qq-$exact.

From the supercharges preserved by the chiral algebra plane there are two (equivalent) choices we can make
\be
\qq_1 = \QQ^1_-  + \zeta \St^{2 \md} \,, \qquad \qq_2 = \SS_1^- - \frac{1}{\zeta} \Qt_{2 \md} \,, 
\ee
where $\zeta$ is an arbitrary phase. The conjugate supercharges are (recall that we work in radial quantization)
\be 
\qq_1^\dagger  = \SS_1^-  +\frac{1}{\zeta} \Qt_{2 \md} \,, \qquad 
\qq_2^\dagger = \QQ^1_-  - \zeta \St^{2 \md} \,.
\ee
The two choices of $\qq_{i=1,2}$ are equivalent since we will see that the cohomology of both families coincide, and are independent of $\zeta$. We could set $\zeta=1$ but will keep it as a bookkeeping parameter.

We find the following $\widehat{sl}(2)=diag(\overline{sl(2)},su(2)_R)$ algebra that is $\qq-$exact 
\bea
\label{eq:Lhatqexact}
-\{\qq_1, \Qt_{1 \md}\} &= \zeta  \{\qq_2, \QQ^2_- \}= \Lh_{-1}\,, \\
-\frac{1}{\zeta}\{\qq_1, \SS_{2}^- \} &= - \{\qq_2, \St^{1\md} \}= \Lh_{1}\,, \\
\{\qq_1, \qq_1^\dagger \} &= \{\qq_2, \qq_2^\dagger \}= \Lh_{0}\,,
\eea
with the generators given by 
\be
 \Lh_{-1}\colonequals \Lb_{-1} - \zeta R_- \,, \qquad \Lh_{+1}\colonequals \Lb_{+1} +\frac{1}{\zeta} R_+\,, \qquad \Lh_0 \colonequals \Lb_0 - \RR\,.
 \ee
This explains the role of $u_I(\zb)$: it was responsible for implementing \emph{twisted} translations, produced by acting with $\Lh_{-1}$ instead of $\Lb_{-1}$, and thus combine a translation with an action of the $su(2)_R$ lowering operator.
Finally the central element $\mathcal{Z}$ is also $\qq-$exact
\be 
\{ \qq_1, \qq_2 \}= -\mathcal{Z}\,.
\ee

We have defined a cohomology where twisted translations by $\widehat{sl}(2)$ are $\qq-$exact, thus if the cohomology at the origin is non-trivial, we can use $\Lh_{-1}$ and $L_{-1}$ to translate operators in $\zb$ and $z$, with the resulting cohomology classes being independent of $\zb$. It remains to be seen that the cohomology at the origin is non-trivial.

We will look simultaneously for the cohomologies at the origin of $\qq_{i=1,2}$ and we shall see they are identical. 
Since both $\Lh_0$ and $\ZZ$ commute with $\qq_i$, and among themselves, we can restrict to operators with a definite eigenvalue under both -- we will look at their eigenspaces.
Furthermore, since they are both $\qq-$exact, an operator in cohomology must have zero eigenvalue under both.
\begin{exercise}
Show this last statement.
\end{exercise}
This requires operators in cohomology to obey
\be 
\Delta - \frac{1}{2}(j +\jb) -R = 0\,, \qquad -r + (j-\jb) =0\,,
\label{eq:Schurcond}
\ee
where $j$ and $\jb$ are Dynkin labels for the two Lorentz spins, and $R$ the Dynkin label for the $su(2)_R$.\footnote{
Note that because we are using Dynkin labels it follows that the $j_{1,2}$ spins of \cite{Beem:2013sza} are related to the Dynkin labels  by $j=2j_1$, $\jb = 2j_2$, and also the eigenvalue $R_{here}=2R_{\text{\cite{Beem:2013sza}}}$.}
Since $\{\qq_i^\dagger,\qq_i \} = \Lh_0$ and the four-dimensional theory is unitary it follows that an operator with zero eigenvalue under $\Lh_0$ is killed by $\qq_{i=1,2}$ and $\qq_{i=1,2}^\dagger$. Note that in defining the cohomologies $\zeta$ played no role.

All in all, \eqref{eq:Schurcond} fully characterizes the harmonic representatives of the $\qq_i$ cohomologies at the origin. We call operators obeying \eqref{eq:Schurcond} \emph{Schur} operators, since they are precisely the ones that contribute to the Schur limit of the superconformal index discussed in \cite{Abhijit}.

This gives a non-trivial cohomology of operators at the origin. We can then translate operators away from the origin by acting with $L_{-1}$ and $\Lh_{-1}$. Recall that
\be
[L_{-1},\phi] = \partial_z \phi \,, \qquad  [\Lb_{-1},\phi] =\partial_{\zb} \phi \,, 
\ee
and operators are translated by exponentiating these actions
\be 
e^{z L_{-1} + \zb \Lb_{-1}} \phi(0,0) e^{-z L_{-1}-\zb \Lb_{-1}} = \phi(z,\zb)\,,
\ee
as can be checked by expanding the exponentials. Then, an operator in cohomology at a position $(z,\zb)$ is obtained by the twisted translations
\be 
e^{z L_{-1} + \zb \Lh_{-1}} \phi(0,0) e^{-z L_{-1}-\zb \Lh_{-1}} =  \phi(z,\zb) - \zb \zeta [\RR_- , \phi(z, \zb)]  +\frac{1}{2} \zeta^2 \zb^2[\RR_- , [\RR_- , \phi(z, \zb)] ]+ \ldots  \,,
\label{eq:twistedtrans}
\ee
where $\zeta$ should not appear in any observable. 
Finally, we note that Schur operators are always the highest weights of the Lorentz and $su(2)_R$ representations. If this was not the case, since Schur operators have $\Lh_0=0$, their respective representations would contain operators with negative $\Lh_0$, in conflict with unitarity which requires $\Lh_0 =\{ \qq_i^\dagger, \qq_i\} \geqslant0$. By going through the list of superconformal representations we also see that non-trivial Schur operators will always transform in non-trivial $su(2)_R$ representations. 

Let us now give an explicit example of twisted translations and recover the $u_I(\zb)$ vector used in the claim~\ref{res:subsector}. Take a Schur operator transforming as a doublet of $su(2)_R$, the operator in cohomology at the origin is the highest weight $\OO^1(0,0)$, and when the operator is translated away from the origin we must also add a lower $su(2)_R$ weight $[\RR_-,\OO^1(0,0)]=\OO^2(0,0)$ according to \eqref{eq:twistedtrans}. This is equivalent to contracting the doublet index of $\OO$ with the vector
\be 
u_i(\zb) = (1 , - \zeta \zb)\,.
\ee
Similarly for a spin $\frac{R}{2}$ operator we would contract the fundamental indices as
\be 
e^{z L_{-1} + \zb \Lh_{-1}} \phi(0,0) e^{-z L_{-1}-\zb \Lh_{-1}} = u_{i_1}(\zb) \ldots u_{i_R}(\zb) \OO^{i_1 \ldots i_R}(z,\zb)\,,
\label{eq:twistedtransfund}
\ee
\begin{exercise}
Show that $\Lh_0=0$ implies $\ZZ=0$ for a unitary SCFT, \ie, the second condition in \eqref{eq:Schurcond} is redundant. Tip use $\{\QQ_-^1, (\QQ_-^1)^\dagger\}$ and $\{\Qt_{2 \md}, (\Qt_{2 \md})^\dagger\}$ to show this.
\end{exercise}

Finally, we can show that at the level of cohomology the OPE is single valued, that is
\be 
\OO_1(z) \OO_2(0) \sim \sum\limits_{\substack{k\\ \text{Schur}}} \frac{\lambda_{12k}}{z^{h_1+h_2-h_k}}\OO_k(0) + \qq_i-\text{exact}\,,
\ee
with $h_1+h_2-h_k$ an integer.
\begin{exercise}
Show this follows from the OPE of twisted translated Schur operators using $2d$ conformal invariance and $su(2)_R$ selection rules.
\end{exercise} 

To summarize, the cohomology classes of $\qq_i$
\be 
\OO(z) \colonequals [\OO(z,\zb)]_{\qq_i}\,,
\ee
have correlation functions that are meromorphic. Moreover, we can compute the $L_0$ weight of the local operators thus defined as
\be 
h= \frac{\Delta}{2} + \frac{j + \jb}{4} = \frac{R+j+\jb}{2} \in \frac{1}{2} \mathbb{Z}_{\geqslant0}\,.
\label{eq:hchiral}
\ee

Note that so far there is no Virasoro element, there is no infinite dimensional conformal symmetry, just a global $sl(2)$.

\begin{example}{Free hypermultiplet\\}
The free hypermultiplet is a $B_1\bar{B}_1[0,0]_1^{(1;0)}$ superconformal multiplet in the notation of \cite{Cordova:2016emh},\footnote{A $\hat{\BB}_{\frac12}$ multiplets in the classification of \cite{Dolan:2002zh}.} and the superconformal primaries are scalars of dimension one, $r=0$ and $su(2)_R$ doublets, as discussed in \cite{Lorenz}:
\be 
q^{i} =  \begin{pmatrix} q \\ \tilde{q}^* \end{pmatrix}\,, \qquad 
\tilde{q}^i =  \begin{pmatrix} \tilde{q} \\ - q^* \end{pmatrix}\,,
\label{eq:freehyper}
\ee
where $i=1,2$ is an $su(2)_R$ index. The highest weights are Schur operators, satisfying \eqref{eq:Schurcond}. The $L_0$ weight of these operators will be
\be 
h= \frac{1}{2}\,,
\ee
which is half-integer even though they are commuting bosons. This is the first time we see a sign of the non-unitarity of the chiral algebras of $\NN=2$ SCFTs.
The superconformal primaries $q(0,0)$ and $\tilde{q}(0,0)$ are in cohomology at the origin, and the twisted translated operators in cohomology are given by \eqref{eq:twistedtransfund}
\be 
\begin{split}
q(z) &\colonequals \left[q(z,\zb) - \zeta \zb \tilde{q}^* (z,\zb) \right]_\qq\,, \\
\tilde{q}(z) &\colonequals \left[\tilde{q}(z,\zb) + \zeta \zb q^* (z,\zb) \right]_\qq\,.
\end{split}
\ee
Let us now take the OPE
\be 
q(z) \tilde{q}(0)  = \underbrace{q(z,\zb) \tilde{q}(0,0)}_{=0} - \zeta \zb \underbrace{\tilde{q}^*(z,\zb) \tilde{q}(0,0)}_{\frac{1}{(z \zb)}} \sim -\zeta \frac{1}{z}\,.
\ee
which is meromorphic and the scaling dimension of $q(z)$ and$\tilde{q}(z)$ comes out $h=\frac12$ as expected.
The singularity of the OPE is controlled by a single term of the $4d$ OPE. This may not be so useful for a free theory but soon we will fix similar correlation functions for strongly coupled theories as well.
All other operators in cohomology are written as normal-ordered products (with derivatives -- derivatives in $z$ direction preserve the Schur condition) of $q(z)$ and $\tilde{q}(z)$, \ie, $(q \ldots \tilde{q} \ldots \partial q \ldots \partial\tilde{q})(z)$, which is not that surprising in free theory, but the chiral algebra will allow us to write down similar expressions for interacting theories as well.
\end{example}

The free hypermultiplet is the first example we saw of an operator in cohomology, and we can check which other operators are there.

\subsubsection*{Virasoro enhancement}
\label{sec:Virenhancement}

The most universal superconformal multiplet, present in any local SCFT, is the one containing the stress tensor. In an $\NN=2$ SCFT it belongs to the $A_2 \bar{A}_2 [0,0]_2^{(0;0)}$ superconformal multiplet in the notation of \cite{Cordova:2016emh} and introduced in \cite{Lorenz}.\footnote{ It is denoted by $\hat{\CC}_{0,(0,0)}$ in \cite{Dolan:2002zh}.} It contains, apart from the stress tensor ($T_{\mu \nu}$), the supersymmetry currents ( $J_{\aa}^{\mu \II}$ and $\bar{J}_{\mu \aad, \II}$,), the $u(1)_r$ current ($j^{u(1)_r}_\mu$) and the $su(2)_R$ current (${j_{\mu \II}}^\JJ$).

From all the operators in this multiplet only the highest weight of the $su(2)_R$ current obeys \eqref{eq:Schurcond}, having $\Delta=4$, $j=\jb=1$, $r=0$ and $R=2$, it gives rise to an operator in cohomology with $sl(2)$ weight $h=2$.
The $su(2)_R$ current will play an important role as it gives rise to the chiral algebra stress tensor, and is responsible for the promised enhancement of the geometric $sl(2)$ on the chiral algebra plane to a full Virasoro symmetry.

The twisted translations of the $su(2)_R$ current are given by
\be \label{eq:2dST}
T(z) \colonequals\left[\kappa u_\II (\zb) u_\JJ(\zb)  j_{+ \pd}^{\II \JJ} (z,\zb)\right]_\qq=\left[\kappa \left(  j_{+ \pd}^{11} (z,\zb)- 2\zb \zeta  j_{+ \pd}^{12} (z,\zb)+ \zb^2 \zeta^2  j_{+ \pd}^{22}(z,\zb) \right)\right]_\qq\,,
\ee
where $\kappa$ is a normalization to be fixed by demanding the canonical normalization for the two-dimensional stress tensor \eqref{eq:STOPE}.

The two-point function of the $su(2)_R$ current is fixed to be\footnote{This differs from the $su(2)_R$ current defined in \cite{Beem:2013sza} by $j_{here}^{\II\JJ}= 2  J_{there}^{\II\JJ}$.}
\be 
\langle j_\mu^{\II \JJ} (x)j_\nu^{k\LL}(0) \rangle =  -\frac{3 c_{4d}}{ \pi^4} \frac{I_{\mu \nu}}{x^6} \epsilon^{k (\II} \epsilon^{\JJ) \LL}\,, \qquad  I_{\mu \nu}(x) = \delta_{\mu \nu} - 2 \frac{x_\mu x_\nu}{x^2}\,,
\label{eq:su2rOPE}
\ee
where supersymmetry fixes the constant in terms of the four-dimensional central charge $c_{4d}$ -- the two-point function of the stress tensor -- see, \eg, \cite{Shapere:2008zf}. For convinience the two-point function of the stress tensor is given in \eqref{eq:STtwopoint}. Here the brackets mean indices are symmetrized and we always take symmetrizations with strength one. The three-point function of conserved currents is given, for example, in eq.~(3.7) and~(3.9) of \cite{Osborn:1993cr}, where Ward identities are used to fix precisely the coefficients of the three-point function in terms of the two-point function in their eq.~(6.12) (noting $C_V$ as defined there is given by $C_V=\frac{3c}{\pi^4}$).
Combining these expressions we find that the OPE of the twisted translated $su(2)_R$ current becomes \cite{Beem:2013sza}
\be 
T(z) T(0) \sim \frac{6 c_{4d} \kappa^2 \zeta^2}{\pi^4} \frac{1}{z^4} - \ii \frac{2 \kappa \zeta}{\pi^2} \frac{T(0)}{z^2} + \qq_i\text{-exact} + \ldots \,.
\ee
\begin{exercise}
Show this.
\end{exercise}
The normalization is then fixed to be
\be 
\kappa = \frac{\ii \pi^2}{\zeta}\,,
\label{eq:kappaST}
\ee
and we obtain the following relation between  the four-dimensional central charge and the two dimensional one\footnote{We take the standard conventions for the central charge in $\NN=2$ SCFTs in which a single free hypermultiplet has $c_{4d}=\frac{1}{12}$ and a single free vector multiplet has $c_{4d}=\frac{1}{6}$.}
\be 
\label{eq:c2dc4d}
c_{2d}=-12 c_{4d} \leqslant 0\,.
\ee
We immediately see that we get a non-unitary chiral algebra, with negative central charge.

To show Virasoro enhancement of the global $sl(2)$ we still need to show that the global symmetry generators ($L_n$ in \eqref{eq:2dsl2}) match the modes of the stress tensor ($\LT_n$) defined by the mode expansion of \eqref{eq:2dST}, \ie, $\LT_{0,\pm1}=L_{0,\pm1}$, when acting on local operators, and such that the scaling weight of the operators under $T(z)$ is given by \eqref{eq:hchiral}. This remains a conjecture in general, and was shown to be the case when $T(z)$ acts on scalar operators, as well as in all known examples \cite{Beem:2013sza}.

\subsection{Properties of the chiral algebra}
\label{sec:props}

Now that we have constructed the map~\ref{res:map} let us look at its properties. When given a particular four-dimensional SCFT how will the chiral algebra we get under the map look like? In all known examples the chiral algebra obtained is strongly finitely generated, meaning we need to understand what are its strong generators, and then all other operators can be written as normal-ordered products (with derivatives) of this finite number of generators. The chiral algebras arising from four-dimensional SCFTs will also have to be very special, inheriting properties from the four-dimensional theory, and in particular four-dimensional unitarity places strong constraints on the allowed chiral algebras -- this will be the topic of the next section. Now let us look at some properties of this map.

\subsubsection*{Independence of exactly marginal couplings}

The chiral algebra  was shown in \cite{Beem:2013sza} to be independent of any exactly marginal deformations in the four-dimensional SCFT. This is achieved by a non-renormalization theorem of 	\cite{Baggio:2012rr} obtained by using superconformal Ward identities. Exactly marginal deformations are the top components of $B_1\bar{L}[0,0]_2^{(0;-4)}$ and their conjugate $L\bar{B_1}[0,0]_2^{(0;4)}$ multiplets. These are the only $\NN=2$ superconformal multiplets that can accommodate exactly marginal deformations, \ie, deformations that preserve supersymmetry -- hence must be killed by all supercharges making them top components -- have dimension four and be neutral under all $R-$symmetries. To show coupling independence of three point functions, $\langle \OO_1(x_1) \OO_2(x_2) \OO_3(x_3) \rangle$, of Schur operators one needs to show that the following four-point function
\be 
\langle \OO_1(x_1) \OO_2(x_2) \OO_3(x_3) \OO_\tau(x) \rangle = 0\,, \qquad \forall x\,,
\ee
when $x_{i=1,2,3}$ are restricted to the chiral algebra plane, and $\OO_\tau$ is the top component of the $B_1\bar{L}[0,0]_2^{(0;-4)}$ superconformal multiplet. Such a non-renormalization theorem follows directly from the results of \cite{Baggio:2012rr} for $\NN=(0,4)$ SCFTs in two dimensions, since we can use a conformal transformation to bring $x$ to the plane, on which the superalgebra we considered is precisely that one.

\subsubsection*{Schur operators are $sl(2)$ primaries}
\label{sec:Schursl2}

We have seen Schur operators are the ones in $\qq-$cohomology, furthermore the Schur operators that are conformal primaries in $4d$, \ie, annihilated by $K_{\mu}$ will automatically be \emph{$sl(2)$ primaries} in chiral algebra, since $L_{+1}=K^{\pd +}$. They will \emph{not}, however, always be Virasoro primaries, as the requirements for a Virasoro primary do not follow from four-dimensional physics. It will be important to keep this distinction in mind and we will see that superconformal symmetry ensures certain representations always give rise to Virasoro primaries, while others can either be Virasoro primaries or descendants. We've already encountered an example of a superconformal multiplet that gives rise to an $sl(2)$-primary that is a Virasoro descendant -- the stress tensor supermultiplet and in particular the $su(2)_R$ current. One can also check that to organize Schur operators in Virasoro representations one is forced to take linear combinations of $2d$ operators of the same dimension, but that arise from \emph{different} four-dimensional superconformal multiplets.

The full list of $\NN=2$ superconformal multiplets  containing Schur operators can be obtained by going through the representations tables of \eg, \cite{Dolan:2002zh}. It consists of \cite{Beem:2013sza}
\be 
B_1 \bar{B}_1\,, \qquad A_{1,2} \bar{B}_1\,, \qquad B_1 \bar{A}_{1,2}\,, \qquad A_{1,2}\bar{A}_{1,2}\,.
\ee
Each of these superconformal multiplets contributes with exactly one Schur operator.

\subsubsection*{A filtration by $su(2)_R$}
\label{sec:filtration}

Schur operators are labeled by three Cartans, which we can take to be $R,j,\jb$, with $\Delta$ and $r$ fixed by \eqref{eq:Schurcond}. However the chiral algebra only preserves a combination of two of these Cartans, namely
\be 
h= \frac{1}{2}(R + j + \jb)\,, \qquad \text{and } \quad -r= \jb - j\,.
\ee
The Cartan of the $su(2)_R$ representation is violated by the chiral algebra, as it is clear by the fact that twisted translations involve operators with different values of the Cartan. The chiral algebra OPE will then violate $R$-charge conservation, but always with negative sign \ie, always by allowing for operators with lower value of $R$, meaning we have a filtration by $su(2)_R$ \cite{Beem:2017ooy}. It has not been understood if it is possible to recover this filtration from a purely chiral algebra point of view. 
As a consequence if we are given a two-dimensional operator, identifying which four-dimensional operator gave rise to it may be ambiguous. We will see examples of this in section ~\ref{sec:consequences}.
The free field realizations of \cite{Bonetti:2018fqz,Beem:2019tfp} for some of these chiral algebras give a canonical proposal for the filtration.

\subsubsection*{Stress tensor supermultiplet}

We have seen in section~\ref{sec:Virenhancement} that the stress tensor supermultiplet contributes with a single $sl(2)$-primary. The four-dimensional $su(2)_R$ current two and three-point functions are fixed by two anomaly coefficients $c_{4d}$ and $a_{4d}$, but only the former appears in correlation function in cohomology.  The stress tensor is never a Virasoro primary, and it can either be a generator of the chiral algebra, or a composite operator made out of normal ordered products of other operators -- we will see examples of both in section~\ref{sec:examples}.

\begin{exercise}{Free hypermultiplet/vector multiplet\\}
Work out the free hypermultiplet example in detail (or free vector multiplet in which case the fermions in the vector multiplet $\lambda_+$, $\tilde{\lambda}_\pd$ described in \cite{Lorenz} will be the Schur operators). What you will find is a $(\beta,\gamma)$ system with weight $(\frac12,\frac12)$ (small $(b,c)$ ghost system of weight $(1,0)$). Write down the twisted translated Schur operators in each of these multiplets, construct the four-dimensional $su(2)_R$ current and check that it matches the $2d$ stress tensor. Check that the OPE between the stress tensor and these operators comes out correctly, finding that $\LT_{0,\pm1}=L_{0,\pm1}$, and that the $2d$ central charge is the predicted value. 
Notice that in this example the stress tensor is not a generator.
\end{exercise}

\subsubsection*{Flavor symmetries}
\label{sec:flavorsymm}

Continuous flavor symmetries of a four-dimensional SCFTs are  continuous symmetries that commute with the superconformal algebra. The conserved current that generates a symmetry is a top component of  the  $B_1 \bar{B}_1 [0,0]_2^{(2;0)}$ half-BPS superconformal multiplet.\footnote{These multiplets are denoted by $\hat{\BB}_1$ in the classification of \cite{Dolan:2002zh}.} 
The flavor current itself is not a Schur operator, however the superprimary of the multiplet is. This corresponds to a dimension two scalar that is a triplet of $su(2)_R$ and, by belonging to the same multiplet of the current, transforms in the adjoint representation of the flavor symmetry. We will denote it by $\mm^{A\, \II\JJ}$, where $\II,\JJ$ are $su(2)_R$ fundamental indices, and $A$ is a flavor adjoint index.
An example of this operator in digression~1.1 of \cite{Mario} (on $su(2)$ superconformal QCD) is the meson operator $\mm_{\ell_2}^{\ell_1}= (\bar{q}^\dagger)^{\ell_1}_A \bar{q}_{\ell_2}$.
The four-dimensional OPE of these operators is\footnote{The conventions in these lectures are different from \cite{Beem:2013sza} since $\mm_{here} = \ii/\sqrt{2} M_{there}$.}
\be
\mm^{A\, \II\JJ}(x)\mm^{B\, k \LL}(0)\sim
 \frac { k_{4d} }{ 32 \pi^4 } \frac{\epsilon^{k(\II}\epsilon^{\JJ)\LL} \delta^{AB}}{x^4}
-\frac{ 1}{ 4 \pi^2 }   \frac{\ii f^{AB}_{\phantom{AB}C}\mm^{C\,(\II(k}\epsilon^{\LL)\JJ)}}{x^2}+\cdots\,,
\label{eq:momentmapope}
\ee
where $A,B,C$ are again adjoint indices, and $f^{AB}_{\phantom{AB}C}$ the structure constants of the algebra.
The coefficients appearing in this OPE are fixed by supersymmetric Ward identities \cite{Dolan:2001tt} in terms of those of the flavor current itself, which we take to have the following two-point function,  following the conventions of \cite{Argyres:2007cn},
\be
 \langle J^A_\mu(x)   J^B_\nu ( 0)  \rangle =  \frac{3 k_{4d}}{4 \pi^4} \delta^{AB} \frac{I_{\mu \nu}}{x^6} \,,
 \label{eq:flavorcurrent}
\ee
here $k_{4d}$ is the central charge associated with the four-dimensional flavor symmetry.\footnote{We use conventions for $k_{4d}$ that are standard for $\NN=2$ SCFTs, see \eg, \cite{Argyres:2007cn}. In these conventions a single free hypermultiplet has an $su(2)$ flavor symmetry with $k_{4d}=1$.}
The twisted translated \eqref{eq:twistedtransfund} $\mm^{A\, \II\JJ}(x)$ reads
\be \label{eq:2dflavor}
\begin{split}
J^A(z) &\colonequals \left[\kappa_J u_\II(\zb) u_\JJ(\zb)  \mm^{A \, \II \JJ} (z,\zb)\right]_\qq\\
&=\left[\kappa_J \left(  \mm^{A \, 11} (z,\zb)- 2\zb \zeta  \mm^{A \, 12} (z,\zb)+ \zb^2 \zeta^2  \mm^{A \, 22}(z,\zb) \right)\right]_\qq\,.
\end{split}
\ee
with \eqref{eq:momentmapope} giving rise to the following OPE in chiral algebra
\be 
J^A(z) J^B(0) \sim  \frac{- k_{4d} \kappa_J^2 \zeta^2 \delta^{AB}}{32 \pi^4 z^2} + \frac{\kappa_J \zeta \, \ii f^{ABC} J^C(0)}{4 \pi^2 \, z}+  \qq \text{-exact}  +\ldots \,.
\label{eq:AKMOPE}
\ee
Fixing
\be 
\kappa_J= \frac{4 \pi^2}{\zeta}\,,
\label{eq:kappaJ}
\ee
this defines the OPE of an Affine Kac Moody (AKM) current algebra with level \cite{Beem:2013sza}\footnote{In our conventions the longest root of the flavor algebra has length $\sqrt{2}$, implying the level of the current algebra ($k_{2d}$) is equal to the two-point function of the AKM currents. See also exercise~\ref{ex:AKM}. }
\be 
k_{2d} = - \frac{1}{2} k_{4d}\,.
\label{eq:k2dk4d}
\ee 
This was one example of a ``Higgs branch'' operator, that made an appearance in \cite{Lorenz,Mario}.

\subsubsection*{ ``Higgs branch''  operators}

The superconformal primaries of $B_1 \bar{B}_1[0;0]_R^{(R;0)}$  multiplets are in $\qq_i-$cohomology and thus captured by the chiral algebra. 
This is the type of multiplet that can accommodate the operators that parameterize the Higgs branch described in \cite{Lorenz,Mario}, and for this reason they are often called ``Higgs branch'' operators. Note, however, that the association of these multiplets with the Higgs branch is conjectural, and it is in principle possible that they appear in a SCFT without corresponding to a flat direction. All our statements here rely only on the superconformal representation of the operator. 

OPE selection rules can be used to show that this type of operators always give rise to \emph{Virasoro} primaries, and moreover, generators of the Higgs branch chiral ring give rise to strong generators of the chiral algebra.

\subsubsection*{Hall-Littlewood operators}

More generally, Hall-Littlewood operators, \ie, operators that contribute to the Hall-Littlewood limit of the superconformal index described in \cite{Abhijit}, are also in cohomology. They are the subset of Schur operators that are in the intersection of two $\NN=1$ chiral rings defined by $\QQ_\alpha^2$ and $\Qt_{2 \ad}$. They are also \emph{Virasoro} primaries, as can be shown by OPE selection rules, and generators of the Hall-Littlewood chiral ring are strong generators of the chiral algebra.

\begin{exercise}
(Hard) Use OPE selection rules: $su(2)_R$ and $u(1)_r$ conservation, plus the Schur condition, to show that Hall-Littlewood operators cannot appear as normal ordered products of non-Hall-Littlewood Schur operators. This establishes that the generators of the Hall-Littlewood ring are strong generators of the chiral algebra. Use OPE selection rules to also show that Hall-Littlewood operators are Virasoro primaries.
\end{exercise}

Notably absent are ``Coulomb branch'' operators, that are not in cohomology and thus play no role in these lectures.

\subsubsection*{Extra supersymmetry}

Throughout the lectures we have been writing $\NN \geqslant 2$ SCFTs since all the statements made here also apply if the theory has more than $\NN=2$ supersymmetry. The construction only requires an $\NN=2$ subalgebra which is the one we have been considering. If the theory in question has $\NN=3$ or $\NN=4$ supersymmetry then some of the extra supercharges commute with $\qq_i$, and the chiral algebra is supersymmetric \cite{Beem:2013sza, Nishinaka:2016hbw,Lemos:2016xke}.
For example, in an $\NN=3$ SCFT the supercharges $\QQ_+^3$ and $\Qt_{3 \pd}$, as well as the corresponding conformal supercharges, commute with $\qq_i$.  Different $sl(2)$ operators in cohomology, coming from different four-dimensional $\NN=2$ superconformal multiplets, are now related by these supercharges, and one finds the following structure
\be 
\begin{tikzcd}
\phantom{\OO_{\mathrm{Schur}}} &\OO_{\mathrm{Schur}} \arrow[dl, "\QQ_+^3" ]\arrow[dr,"\Qt_{3 \pd}"] &  \phantom{\OO_{\mathrm{Schur}}}\\
\OO'_{\mathrm{Schur}}  \arrow[rd,"\Qt_{3 \pd}"]&\phantom{\OO_{\mathrm{Schur}} } &\OO''_{\mathrm{Schur}}\arrow[ld,"\QQ_+^3"] \\
&\OO'''_{\mathrm{Schur}} &
\end{tikzcd}
\ee

\begin{exercise}
Find the supercharges of $\NN=3,4$ that commute with $\qq_i$ and find their superalgebra in $2d$ -- you will recover the $\NN=2$ and small $\NN=4$ superalgebras respectively. Note: you will see that part of the extra R-symmetry generators also commute with $\qq_i$ and are thus part of the $2d$ algebra. 
\end{exercise}

\subsubsection*{Superconformal index}
\label{sec:index}
The chiral algebra preserves the four-dimensional Cartans $L_0$ and $-r=\jb-j$, and so we define a (graded) partition function\footnote{Note that we factored out an overall power of $q^{-c_{2d}/24}$. This normalization is typically included in chiral algebra partition functions, and it must be added for the modular properties of \cite{Beem:2017ooy} to hold.}
\be 
Z(q,x) = \Tr\left( q^{L_0} x^{\jb-j}  \right) = \Tr\left( q^{\Delta - R} x^{F}\right)\,,
\label{eq:partfunct}
\ee
where $F=\jb-j$ is the fermion number.
Taking $x=-1$ we recover exactly the Schur limit of the superconformal index, $\mathcal{I}(q)$, introduced in \cite{Abhijit}. Note that Schur operators are also the ones contributing in the Macdonald limit of the superconformal index, which has an extra fugacity, $t$, that keeps track of an additional Cartan. However, the $R$ symmetry grading is lost in chiral algebra as discussed above in section~\ref{sec:filtration},  and thus the Macdonald index has no direct counterpart in chiral algebra.\footnote{For proposals on how to recover the Macdonald index in chiral algebra see \cite{Song:2016yfd,Bonetti:2018fqz,Beem:2019tfp,Beem:2019snk,Xie:2019zlb}.}

So far only the $c_{4d}$ anomaly coefficient as made an appearance in the chiral algebra, however the $a_{4d}$ Weyl anomaly can be recovered  studying the $q\to 1$ limit of the superconformal index. It has been suggested \cite{Buican:2015ina, DPKR, Ardehali:2015bla}, generalizing the arguments of \cite{DiPietro:2014bca} that in this limit the index behaves as
\be 
\lim_{q \to 1} \mathcal{I}(q) \sim e^{\frac{8 \pi^2}{\beta} \left(c_{4d}-a_{4d}\right)} \,,
\ee
where $q \equalscolon e^{-\beta}$.

\subsection{Examples}
\label{sec:examples}

If we are given a four-dimensional SCFT, we would now like to find its associated chiral algebra. As we have seen Higgs branch (and Hall-Littlewood)  chiral ring generators will immediately be strong generators of the chiral algebra. Depending on the theory the stress tensor may, or may not, be a generator. This makes up a reasonable guess for the chiral algebra, and an analysis of the superconformal index can give hints at additional generators.\footnote{Note that due to the fact that the index counts operators with signs there can be cancellations and thus it is not always clear what the set of generators are.} One can then write down the most general singular OPEs between these operators and impose associativity of the operator product algebra \eqref{eq:assoc}. A nice mathematica package for chiral algebras makes computations much simpler \cite{Thielemans:1991uw,Krivonos:1995bk}.
The chiral algebras of a large set of theories has been constructed in this manner, see \eg, \cite{Lemos:2014lua,Nishinaka:2016hbw,Choi:2017nur,Beem:2019snk}.

\subsubsection*{Argyres Douglas SCFTs}
\label{sec:ADth}

A sequence of Argyres-Douglas SCFTs, $(A_1,A_{2n})$, has been conjectured \cite{rastelli_harvard,Beem:2017ooy} to have as chiral algebras the non-unitary $(2,2n+3)$ Virasoro minimal models. This conjecture has been checked by a matching of central charges, and the superconformal index \cite{Cordova:2015nma,Buican:2015ina}. The simplest example, $(A_1,A_2)$, corresponds to the ``simplest'' Argyres-Douglas SCFT, found in the original work of \cite{Argyres:1995jj} on the Coulomb branch of pure $su(3)$ gauge theory. This is an example of an Argyres-Douglas theory \cite{Argyres:1995jj,Argyres:1995xn} described in \cite{Mario}. Its chiral algebra corresponds to the Lee-Yang minimal model. All of these theories are strongly coupled isolated fixed points, \ie, they have no exactly marginal deformations, rendering standard Lagrangian techniques ineffective.\footnote{In \cite{Maruyoshi:2016tqk,Maruyoshi:2016aim,Agarwal:2016pjo,Agarwal:2017roi,Benvenuti:2017bpg} $\NN=1$ Lagrangians that flow to some of these Argyres-Douglas theories were obtained.}
In chiral algebra, however, they are very simple, and \emph{all} Schur operators are normal ordered products (with derivatives) of the stress tensor.

\subsubsection*{$\NN=2$ Superconformal QCD}

Let us now consider a Lagrangian example, superconformal QCD, \ie, a theory with $su(N)$ gauge group and $N_f=2N$ fundamental hypermultiplets.
The chiral algebras of these theories can be constructed from the free theory ones, by taking the Schur operators in the free hypermultiplets and vector multiplets
\be 
q^{i}_{f=1,\ldots,N_f \, a=1,\ldots,N}\,, \qquad \tilde{q}^{i  \,  a=1,\ldots N \, f=1,\ldots N_f}\,, \qquad 
\lambda^A_+\,, \qquad \tilde{\lambda}_{\pd A}\,,\
\ee
where $i$ is an $su(2)_R$ fundamental index and $A$ an $su(N)$ adjoint index and $f$ a flavor index. Note that with respect to \eqref{eq:freehyper} we now have that $q^i$ transforms in the fundamental of $su(N_f)$ and of $su(N)$, while $\tilde{q}$ transforms in the conjugate representations.
Starting from the free ingredients  one can obtain the chiral algebra of the interacting theory by performing the chiral algebra image of four-dimensional gauging. This prescription was put forward in \cite{Beem:2013sza} and corresponds to restricting to gauge invariant operators and performing a certain BRST-cohomological computation that removes the short multiplets that recombine to form longs as the gauge coupling is turned on. As couplings are turned on operators can acquire anomalous dimensions. As discussed in exercise~4 of \cite{Lorenz}, and in \cite{Abhijit}, the only way a short operator can acquire an anomalous dimension  is to recombine with other shorts to form a long multiplet, whose dimension is no longer fixed and can be a non-trivial function of the gauge coupling. This gauging  was applied  to these theories in \cite{Beem:2013sza}, and the  resulting low dimensional spectrum of the chiral algebra was obtained. However this procedure becomes rather cumbersome easily.

The second option to obtain these theories is to guess what are the strong generators and find an associative operator product algebra. Let us consider the case of an $su(2)$ gauge group. Since the fundamental of $su(2)$ is pseudoreal  we have that with $N_f=4$ the flavor symmetry of the theory is enhanced to $so(8)$ as discussed in digression~0.1 and digression~1.1 of \cite{Mario}. The minimal guess of a chiral algebra turns out to be sufficient, it corresponds to an $\widehat{so}(8)_{-2}$ affine Kac-Moody current algebra, with level $k_{2d}= - \frac{1}{2}k_{4d} = -2$. This guess is motivated by the flavor symmetry and by the fact that the Higgs branch of the theory is generated simply by the superprimaries of the $so(8)$ flavor currents. In this case the stress tensor is not an independent generator, as it can be constructed from the normal ordered product of two currents via the Sugawara construction
\be 
T(z) = \NN_T (J^A J^B)(z)\,,
\ee
with $\NN_T$ a normalization fixed by demanding the canonical OPE for $T(z)$ \eqref{eq:STOPE}. One can compute the resulting central charge and one finds $c_{2d}=-14$. From \eqref{eq:c2dc4d} we see this is in agreement with the central charge of $su(2)$ SQCD of $c_{4d}=\tfrac{7}{6}$. As such there is no need to add $T(z)$ as an extra generator, and one can check that the superconformal index matches the vacuum character of the theory. Furthermore, it can be checked that the low dimensional operators of the resulting chiral algebra match the ones obtained through the gauging procedure \cite{Beem:2013sza}.

\subsubsection*{Class $\mathcal{S}$}

As described in \cite{Bruno} the elementary building blocks in class $\SS$ are trinions, \ie, three punctured spheres with maximal punctures. One can then obtain any class $\SS$ theory by a combination of two operations:
\begin{itemize}
\item Gauging of two flavor symmetries: the chiral algebra counterpart of this corresponds to the aforementioned gauging prescription developed in \cite{Beem:2013sza}.
\item Reducing a flavor symmetry of a puncture: the chiral algebra procedure corresponding to this action was put forward in \cite{Beem:2014rza} as a quantum Drinfeld-Sokolov reduction \cite{Feigin:1990pn,deBoer:1993iz}.
\end{itemize}
Note, however, that both of these procedures involve computing cohomologies and are technically very cumbersome.
Proposals for the chiral algebras of trinion theories have appeared in  \cite{Beem:2014rza, Lemos:2014lua,Arakawa:2018egx,Beem:2020pry}.
In an enlarged class $\SS$, where one also adds irregular punctures, the building blocks will also include spheres with one maximal and one irregular puncture.
Dual descriptions of the same theory in class $\SS$ now give predictions for relations between chiral algebras.

%% file: sections/4_consequences.tex
\section{Consequences for four-dimensional physics}
\label{sec:consequences}

So far we looked at the map of \ref{res:map} in one direction, exploring how different features of four-dimensional SCFTs manifest themselves in chiral algebra, and seeing how to obtain the chiral algebra associated to a given SCFT.
The chiral algebras can often be obtained by starting from the expected strong generators and fixing their singular OPEs by demanding associativity of the operator product algebra \eqref{eq:assoc}. Fixing the singular OPEs amounts to fixing CFT data of the parent four-dimensional SCFT. Moreover, one can then compute OPE coefficients of normal ordered products (including derivatives) of the generators from the singular part, in principle fixing an infinite amount of CFT data in four-dimensions. The only difficulty that can arise resides in identifying which four-dimensional operators, \ie, superconformal representation, corresponds to a given $2d$ operator, which is tied to the lost grading of the $su(2)_R$ Cartan. In some cases such ambiguities will be easy to resolve, while in others one would require knowledge additional information. 

Furthermore, by making only \emph{general assumptions} about the operator content of four-dimensional SCFTs, \eg, local theories, with a given flavor symmetry, or higher supersymmetry, we can fix a \emph{sub-sector common to the chiral algebras of all $4d$ SCFTs that share that property}. In turn, knowledge of the chiral algebra sub-sector can be translated into knowledge of a  sub-sector of the \emph{protected spectrum and OPE coefficients} of the four-dimensional theory.

In fixing these sub-sectors we have not imposed any conditions arising from four-dimensional unitarity.
While the chiral algebras of $4d$ SCFTs are non-unitary, unitarity was broken in a very specific way: some local operators can acquire negative norms. Moreover, whether an operator acquires a negative norm is completely determined from its quantum numbers, so ultimately from which superconformal representation it came from. Recall that unitarity, or reflection positivity, in the four-dimensional theory requires norms of states to be positive.
As we will see the sub-sectors obtained do not automatically satisfy these conditions, leading to new \emph{unitarity bounds} \cite{Beem:2013sza,Beem:2013qxa,Beem:2016wfs,Liendo:2015ofa,
Lemos:2015orc,Cornagliotto:2017dup,Beem:2018duj}.

\subsection{An example of a \texorpdfstring{$4d$}{4d} unitarity bound from chiral algebra}

Consider a four-dimensional SCFT with a flavor symmetry algebra $g_f$. If the $4d$ theory is local then it must have a four-dimensional stress tensor supermultiplet, and thus a two-dimensional stress tensor according to section~\ref{sec:Virenhancement}
\be
 T(z) T(0) \sim \frac{c_{2d}}{z^4} + \frac{2 T(0)}{z} + \frac{\partial T(0)}{z}+ \ldots\,,
 \label{eq:TTOPE}
\ee
which we notice acquired a negative norm.
Furthermore, flavor symmetries give rise to currents as shown in section~\ref{sec:flavorsymm},
\be 
J^A(z) J^B(0) \sim \frac{k_{2d}}{z^2} + \frac{\ii f^{AB}_{\phantom{AB}C} J^C(0)}{z}+\ldots\,,
\label{eq:JJOPE}
\ee
that also acquired a negative norm, and which are Virasoro primaries, \ie,
\be 
T(z) J^A(0) \sim \frac{J^A(0)}{z^2} + \frac{\partial J^A(0)}{z}+\ldots\,.
\label{eq:TJOPE}
\ee
So far this all follows from the four-dimensional SCFT, and we've just summarized the results from the previous section.

The novelty comes in chiral algebra, where  the singular pieces of the known OPEs written above are enough to compute correlation functions of the remaining operators. For example, we can consider the normal ordered product of two currents,
\be 
\label{eq:JJno}
(J^A J^B)(0) = \lim\limits_{z \to 0} \left(J^A(z) J^B(0) - \text{singular terms}\right)\,,
\ee
and compute its correlation functions. From four-dimensions we would not have known if this operator was present, but in chiral algebra that is easy to establish.

Let us consider the singlet piece of the normal ordered product \eqref{eq:JJno}, this is the so called Sugawara stress tensor,
\be 
S(z) = (J^A J^A)(z) \,,
\ee
can we determine to which operator in $4d$ it corresponds to? As described in section~\ref{sec:filtration} identifying the four-dimensional origin of a chiral algebra operator is ambiguous. However, in this case we will be able to settle the ambiguities.
There are only two superconformal multiplets that have, in chiral algebra, the same quantum numbers as $S(z)$ ($h=0$ and $r=0$) and that can appear in the self-OPE of two flavor current supermultiplets: the stress tensor and a Higgs branch operator of dimension four ($B_1 \bar{B}_1[0,0]_4^{(4;0)}$) \cite{Dolan:2001tt,Nirschl:2004pa,Dolan:2004mu}. Let us call the chiral algebra image of the latter $B(z)$, then it follows that
\be 
S(z) = \beta_1 T(z) + B(z)\,,
\ee
where we are not picking any specific normalization for $B(z)$.
While these two operators appear degenerate in chiral algebra, they differ in four-dimensions, and thus their two-point function will vanish. This allows us to obtain
\be 
\beta_1 = \langle T(z) S(0) \rangle  z^4 \frac{c_{2d}}{2} = 2 \frac{\dim_{g_f} k_{2d}}{c_{2d}}\,,
\label{eq:beta1}
\ee
thus getting 
\be 
B(z) = S(z) - 2 \frac{\dim_{g_f} k_{2d}}{c_{2d}} T(z)\,.
\ee
Note that fixing \eqref{eq:beta1} is one example of computing correlation functions of normal-ordered products from the singular OPEs, in particular we get from \eqref{eq:JJOPE} and \eqref{eq:TJOPE}
\be 
\langle T(z) J^A(z_1) J^B(0) \rangle = \frac{\delta^{AB} k_{2d}}{(z-z_1)^2 z^2}\,,
\ee
from which \eqref{eq:beta1} follows by taking the limit $z_1 \to 0$ to obtain the normal ordered product \eqref{eq:JJno}.

Thanks to the chiral algebra we can now compute correlation functions involving the  $B(z)$ Higgs branch operator, which would not have been possible from a purely four-dimensional perspective. In particular we can compute its two-point function, \ie, its norm, which we know from four-dimensional unitarity must be positive. 
The two-point function of the Sugawara stress tensor is given by (see example~\ref{exa:sug})
\be 
\langle S(z)  S(0) \rangle = \frac{2 \dim_{g_f} k_{2d}(k_{2d} + h^\vee )}{z^4}\,,
\label{eq:TsugTsug}
\ee
where $h^\vee$ is the dual Coxeter number, given by $h^\vee =\frac{f^{AB}_{\phantom{AB}C} f_{AB}^{\phantom{AB}C}}{2 \dim_{g_f}}$\footnote{We are using conventions where the length of the longest root of $g_f$ is $\sqrt{2}$}.
Finally we get
\be 
\langle B(z) B(0) \rangle = 2 \dim_{g_f} k_{2d}^2 \left(1 + \frac{h^\vee}{k_{2d}}- \frac{\dim_{g_f}}{c_{2d}}\right)\,,
\ee
and imposing positivity of the four-dimensional norm, we get a new unitarity bound \cite{Beem:2013sza}
\be 
\frac{\dim_{g_f}}{c_{4d}} \geqslant \frac{24 h^\vee}{k_{4d}} -12\,,
\label{eq:c4dk4dbound1}
\ee
where we already used the maps to the four-dimensional central charges \eqref{eq:c2dc4d} and \eqref{eq:k2dk4d}.
This is a new unitarity bound on the four-dimensional central charges, that only exists because we could compute the two-point function of a particular $B_1 \bar{B}_1[0,0]_4^{(4;0)}$ operator for any local theory with a flavor symmetry from the chiral algebra.
Furthermore, we see that the norm of this operator goes to zero, \ie, it becomes null, and the Sugawara matches the stress tensor, when the inequality \eqref{eq:c4dk4dbound1} is saturated. This implies four-dimensional SCFTs saturating this bound have a particular relation on their Higgs branch chiral ring, setting 
\be
\lim\limits_{z\to 0} M^{11 A} (z) M^{11 A}(0)  = 0\,,
\label{eq:chiralringrel}
\ee
where $11$ means we are taking the $su(2)_R$ highest weight, and where we took the singlet term of the chiral ring. Recall this OPE is non-singular, hence giving rise to a chiral ring as shown in \cite{Lorenz}.
Let us look again at digression~1.1 of \cite{Mario}, namely superconformal QCD with gauge group $su(2)$, which has $c_{4d}= \frac{7}{6}$ and $k_{4d}=4$, precisely saturating the bound \eqref{eq:c4dk4dbound1} (note that $h^\vee=6$ and $\dim_{g_f}=28$). The predicted chiral ring relation \eqref{eq:chiralringrel} is precisely the second equation in eq.~(1.12) of \cite{Mario}, obtained there from the F-term conditions.

\begin{example}
\label{exa:sug}
As an example of how to compute correlation functions of normal-ordered operators in chiral algebras from the singular OPEs let us compute the two-point function 
\be 
\langle S(z)  S(0) \rangle = \lim\limits_{\substack{z_1\to z,\\ z_3 \to 0}} \langle
J^A (z_1) J^A(z) J^B(z_3) J^C(0)\rangle - \text{singular terms}\,,
\ee
We start by using the OPE \eqref{eq:JJOPE} to fix the following correlator from its singularities as
\be 
\begin{split}
 & \langle J^A (z_1) J^A(z_2) J^B(z_3) J^B(0)\rangle= \frac{k_{2d}^2 \delta^{AA}\delta^{BB}}{z_{12}^2 z_3^2} + \frac{\delta^{AB} \delta^{AB} k_{2d}^2}{z_{13}^2 z_2^2} + \frac{\ii f^{AAC}}{z_{12}} \langle J^C(z_2) J^B(z_3) J^B(0) \rangle \\ 
&+ \frac{\ii f^{ABC} }{z_{13}} \langle J^A(z_2) J^C(z_3) J^B(0) \rangle + \frac{\delta^{AB} \delta^{AB} k_{2d}^2}{z_1^2 z_{23}^2} + \frac{\ii f^{ABC}}{z_{2}} \langle J^A(z_1) J^B(z_3) J^C(0) \rangle\,,
\end{split}
\ee
where $f^{AAC}=0$ due to anti-symmetry of the structure constants, and where the three-point functions are trivially fixed from \eqref{eq:JJOPE}.
Taking the limits $z_1 \to z_2$ and $z_3 \to 0$ after subtracting the singular pieces yields \eqref{eq:TsugTsug}.
\end{example}

\subsection{Unitarity bounds}

One can get similar bounds by looking at other representations in the normal-ordered product \eqref{eq:JJno}. Selection rules still require this operator to be a linear combination of the  $B_1 \bar{B}_1[0;0]_4^{(4;0)}$ Higgs branch operator and a stress tensor supermultiplet. However, in an \emph{interacting} theory, there will be no stress tensor in non-singlet channels and thus the normal-ordered product is simply the Higgs branch operator, whose two-point function we can compute and require to be positive. This is more efficiently implemented by computing the four-point function of $J^A(z)$ and decomposing it in \emph{$sl(2)$ blocks} \eqref{eq:confblockdec}, such that we extract directly the contribution of $sl(2)$ primaries appearing in the self-OPE of $J^A (z) J^B(0)$, and do not have to subtract descendants.\footnote{In the example above the descendants that could appear at that dimension were $\partial J^A(z)$ which is not a flavor symmetry singlet. Note that we are using  decomposition in $sl(2)$ blocks and \emph{not} Virasoro blocks since as discussed in section~\ref{sec:Schursl2} Virasoro representations generically mix different four-dimensional operators.} Doing so we get the bounds of \cite{Beem:2013sza} listed in table~\ref{tab:bounds}, where we also list the representation that gave rise to the bound. These bounds allow one to determine when Higgs branch chiral ring relations can occur in the product of two $\mm^{11 A}$, see \cite{Beem:2013sza} for an extensive discussion. 

Similarly one can look at the norms of higher dimensional operators, but no new constraints arise.\footnote{The exception is the case of $g=su(2)$, but in that bound will be surpassed by \eqref{eq:newbound} so we will not write it down.} In doing so there will again be ambiguities, which are always resolvable. Contributions of $A_1 \bar{A}_1 [j;j]^{(0,0)}_{2+j}$ and $A_\ell \bar{A}_\ell [j;j]^{(2,0)}_{2+2+j}$ multiplets appear ambiguous, but the former contains conserved currents of spin greater than two if $j >0$. Such currents are absent in interacting theories \cite{Maldacena:2011jn,Alba:2013yda}, and thus imposing the absence of these multiplets we resolve all ambiguities.

\begin{table}[t] 
\centering
\begin{tabular}{lllc}
\hline\hline
$g_f$ &~~~~~~~~~~~~~~~~~~~~& Bound~~~~~~~~~~~~~~~~ & Representation of  $B_1 \bar{B}_1[0;0]_4^{(4;0)}$\\[0.5ex] 
\hline 
$su(N)$ & $N \geqslant 3$ 		& $k_{4d}\geqslant N$				& $\mathbf{N^2-1}_{\mathrm{symm}}$			\\[.3ex]
$so(N)$ & $N = 4,\ldots,8$	& $k_{4d}\geqslant 4$ 	  	 		& $\mathbf{\frac{1}{24} N(N-1)(N-2)(N-3)}$	\\[.3ex]
$so(N)$ & $N \geqslant 8$ 		& $k_{4d}\geqslant N-4$ 				& $\mathbf{\frac12 (N+2)(N-1)}$				\\[.3ex]
$usp(2N)$& $N \geqslant 3$ 		& $k_{4d}\geqslant N+2$				& $\mathbf{\frac12 (2N+1)(2N-2)}$				\\[.3ex]
$g_2$ 	 &					& $k_{4d}\geqslant \frac{10}{3}$  	& $\mathbf{27}$								\\[.3ex]
$f_4$ 	 &					& $k_{4d}\geqslant 5$  				& $\mathbf{324}$							\\[1ex]
$e_6$ 	 &					& $k_{4d}\geqslant 6$           	 	& $\mathbf{650}$							\\[.3ex]
$e_7$ 	 &					& $k_{4d}\geqslant 8$  				& $\mathbf{1539}$							\\[.3ex]
$e_8$ 	 &					& $k_{4d}\geqslant 12$  				& $\mathbf{3875}$							\\[.3ex]
\hline
\end{tabular} 
\caption{Unitarity bounds for the anomaly coefficient $k_{4d}$ arising from positivity of the $B_1 \bar{B}_1[0;0]_4^{(4;0)}$ norm in the non-singlet channels.\label{tab:bounds} }
\end{table} 

So far we have been looking excusively at operators appearing in the OPE of flavor currents. Further constraints can be obtained by lifting degeneracies in four dimensions. There can be more than one operator in a given superconformal representation, and they will couple differently to different four-dimensional operators. As such, by considering simultaneously the OPE \eqref{eq:TTOPE} and \eqref{eq:JJOPE} one can obtain stronger constraints on the space of theories \cite{Lemos:2015orc}, namely:
\be
\label{eq:newbound}
k_{4d} \left(-180 c_{4d}^2+66 c_{4d}+3 \dim_{g_f} \right)+60 c_{4d}^2 h^\vee-22 c_{4d} h^\vee\leqslant 0\,. 
\ee
Further constraints can be obtained by assuming the theory has a reductive flavor group \cite{Beem:2017ooy,Beem:2018duj}. Superconformal QCD with gauge group $su(2)$ also saturates this bound, making it the interacting theory with the smallest possible $k_{4d}$, which fixes $c_{4d}$ uniquely. See \cite{Beem:2013sza,Lemos:2015orc} for a discussion on other cases.

Assuming only the existence of a stress tensor, \ie, considering only \eqref{eq:TTOPE} one finds \cite{Liendo:2015ofa}
\be 
c_{4d}  \geqslant \frac{11}{30}\,,
\label{eq:c4d1130}
\ee
which is saturated by the $(A_1,A_2)$ Argyres-Douglas theory of section~\ref{sec:ADth}, whose chiral algebra is the Lee-Yang minimal model.

Finally, for theories with $\NN=3$ or $\NN=4$ supersymmetry one obtains \cite{Beem:2013qxa,Beem:2016wfs,Cornagliotto:2017dup}
\be 
c_{4d} =a_{4d} > \frac{13}{24} \,, \quad \text{for }\NN=3\,,\qquad c_{4d} =a_{4d} \geqslant \frac{3}{4} \,, \; \text{for }\NN=4\,, 
\ee
where we used that $c_{4d}=a_{4d}$ when we have $\NN \geqslant3$. Recall that all these bounds apply for \emph{interacting} theories, as the assumption of the absence of higher spin currents is necessary to resolve ambiguities. The $\NN=4$ bound of \cite{Beem:2013qxa,Beem:2016wfs} is saturated by $\NN=4$ Super-Yang-Mills with gauge group $su(2)$, while the $\NN=3$ bound of \cite{Cornagliotto:2017dup} cannot be saturated by any interacting theory. This statement follows from showing that with $c=\tfrac{13}{24}$ a two-dimensional operator has the wrong norm to be interpreted as the expected four-dimensional superconformal multiplet (assuming the theory is interacting and there are no conserved currents of higher spin).

\bigskip

To summarize, the chiral algebras arising from four-dimensional SCFTs are very constrained by $4d$ unitarity. These constraints, allied with the rigid structure of chiral algebras and meromorphicity, give rise to a large number of constraints on the allowed space of $4d$ SCFTs. One could also hope to use these constraints to make progress in a classification of all $4d$ $\NN \geqslant 2$ SCFTs. For example, assume a theory whose only generator is the stress tensor, can we check all norms of normal-ordered operators (and not just those appearing in the $T(z)T(0)$ OPE considered above) have the right sign? While in principle this would put constraints on the allowed values of $c_{4d}$ for theories whose only generator is $T(z)$, the ambiguities due to the loss of the $R-$symmetry grading prevent one from naively identifying the four-dimensional origin of the chiral algebra operators.\footnote{We thank B.~van~Rees for many discussions on this.}

%% file: sections/5_outlook.tex

\section{Outlook}
\label{sec:outlook}

We have constructed a map from four-dimensional SCFTs to two-dimensional chiral algebras that is the subject of various directions of ongoing work. Some implications and consequences are discussed in these lectures but it is impossible to do justice to all of the work done on the subject. In this section we'll briefly summarize a few of the general current directions and their implications. The classification of $\NN\geqslant 2$ SCFTs remains an important unsolved problem, and several current directions aim at making progress towards this ambitious goal.


\subsection{Recovering the Higgs branch from the chiral algebra}

We have seen during these lectures that Higgs branch operators are in the $\qq_i$-cohomolgy, and the Higgs branch chiral ring has played a promient role, with its generators appearing as generators of the chiral algebra, and with chiral ring relations showing up as the saturation of unitarity bounds. However, there are many other four-dimensinal operators appearing in chiral algebra that are not part of the Higgs branch chiral ring, notably the stress tensor. This leaves the question if one can distill, in chiral algebra, which operators arose from four-dimensional Higgs branch operators, and whether  one can recover the Higgs branch from the chiral algebra. This question was answered in  \cite{Beem:2017ooy}, where the authors conjectured the Higgs branch chiral ring is obtained from  Zhu's $C_2$ algebra \cite{ZhuThesis} (which roughly speaking means removing from the chiral algebra all normal-ordered products that include derivatives) after removing all nilpotent elements in this algebra (for example the stress tensor had not been removed by the previous step, and so it must be that the $2d$ stress tensor is nilpotent Zhu's $C_2$ algebra). The authors also conjectured that the Higgs branch is equivalent to the the associated variety of the chiral algebra as introduced by Arakawa \cite{arakawa2010remark}. These conjectures were checked in a variety of examples. An immediate consequence of their conjectures is that the Schur index, \ie, the graded vacuum partition function of the chiral algebra, must satisfy a finite order linear modular differential equation. The other solutions correspond conjecturally to indices of the theory in the presence of $\NN=(2,2)$ surface defects. The understanding of the modular behavior of the superconformal index makes contact with the $a_{4d}$ anomaly coefficient that appears in the $q \to 1$ limit of the index as discussed in \ref{sec:index}. The $q \to 1$ limit of the vacuum module is mapped under modular transformations to the small $q$ behavior of a non-vacuum module whose dimension encodes $a_{4d}$.

\subsection{Recovering the chiral algebra from the Higgs branch}

One can now ask the converse question, given the Higgs branch of a four-dimensional $\NN\geqslant 2$ SCFT can we write down the chiral algebra of the SCFT? The first steps in addressing this question were taken in \cite{Beem:2019tfp} (see also \cite{Bonetti:2018fqz} for the case of $\NN=3$ and $\NN=4$ SCFTs), where the authors constructed the chiral algebra of a number of four-dimensional SCFTs. 
In the examples considered the chiral algebras are obtained by considering the effective field theory description of the theory in question on a generic point of the Higgs branch. 
Accordingly, if the generic point of the Higgs branch contains a collection of free hypermultiplets and free vector multiplets, the chiral algebra of the SCFT, which sits at the origin of the Higgs branch, involves a free field realization in terms of free chiral bosons and symplectic fermion pairs.  In cases where the generic point of the Higgs branch has a decoupled interacting SCFT in the infrared, in addition to the free fields, one must also add the chiral algebra of this interacting infrared SCFT as a building block. Note that since one is considering a generic point on the Higgs branch this SCFT must be a theory with a trivial Higgs branch.
This approach was further developed  in \cite{Beem:2019snk} where the authors considered instead a non-generic, more symmetric, locus on the Higgs branch. Starting from the chiral algebra of the SCFTs obtained at this locus the authors find a more economical free field realization of the SCFT.

This picture suggests that the chiral algebras arising from four-dimensional SCFTs can be recovered from the Higgs branch, if one is given the effective field theory description on the Higgs branch.

\subsection{Other dimensions}

The same construction can be applied whenever the superconformal algebra has a $psu(1,1|2)$ subalgebra that is the supersymmetrization of M\"{o}bius transformations on the plane \cite{Beem:2014kka}. Going through the superconformal algebras existing in various dimensions one finds the following list  \cite{Beem:2014kka} that gives rise to chiral algebras\footnote{A similar cohomological construction exists in three-dimensional $\NN\geqslant 4$ SCFTs, leading to a topological theory for the operators restricted to a line \cite{Chester:2014mea,Beem:2016cbd}. The question of obtaining this construction from the chiral algebra discussed in these lectures was addressed in \cite{Dedushenko:2019mzv,Pan:2019shz,Dedushenko:2019mnd}. }
\begin{itemize}
\item Six dimensional $\NN=(2,0)$ SCFTs -- whose superalgebra is $osp(8^\star|4)$,
\item Four dimensional $\NN \geqslant 2$ SCFTs -- whose superalgebra is $su(2,2|\NN\geqslant2)$ and the focus of these lectures,
\item Two-dimensional SCFTs with the ``small'' $\NN=(0,4)$ superconformal algebra, or $\NN=(4,4)$ -- $psu(1,1|2)$
\end{itemize}
Note that the first algebra in two-dimensions is precisely the one that makes an appearance as the chiral algebra of $4d$ $\NN=4$ SCFTs. Being a chiral algebra to begin with the further twist makes the theory position independent. This is a consequence of the larger symmetry of four-dimensional $\NN=4$ SCFTs, that  allows for bigger twists as discussed in \cite{Drukker:2009sf,Liendo:2016ymz}.

The six dimensional case was the subject of \cite{Beem:2014kka}, and likely has implications for a microscopic understanding of the AGT correspondence covered in \cite{Bruno}.
The six dimensional theory also admits codimension two defects, whose worldvolume preserve an $su(2,2|2)$ superconformal algebra with the construction of this lectures directly applying. The chiral algebra of these defects was also studied in \cite{Beem:2014kka}. Note that the four-dimensional worldvolume is of a defect inside a six-dimensional SCFT, and as such the four-dimensional theory will not have a stress tensor since it exchanges with the bulk.
%

\subsection{Enlarging our set of observables -- adding non-local operators}
\label{sec:defects}
The discussion so far has only focused on local observables of the SCFT, but the chiral algebra can be enriched by adding surface defects as shown in \cite{defectLCW,Cordova:2017mhb}.
 The supercharges $\qq_i$ used to define the cohomology are preserved when inserting an  $\NN=(2,2)$ surface defect orthogonal to the chiral algebra plane.
The insertion of the defect gives rise, in cohomology, to non-vacuum modules of the original chiral algebra without defect insertions \cite{defectLCW,Cordova:2017mhb}.
The Schur limit of the superconformal index in the presence of defects matches the  (graded) partition functions of these modules. In \cite{defectLCW,Cordova:2017mhb,Nishinaka:2018zwq} this fact was used to study chiral algebras in the presence of defects, and to propose the chiral algebra version of four-dimension constructions of defects. Turning to correlation functions these can now depend on marginal deformations, since the argument sketched above does not hold in the less symmetric case of having a defect insertion.
A localization computation was set up in \cite{Pan:2017zie}  to compute correlators of Schur operators in the presence of these defects, however, the final expressions could not obtained.
In \cite{Bianchi:2019sxz} properties of the state inserted by the defect identity in chiral algebra were determined from OPE selection rules. In particular, the authors obtained the action of the stress tensor modes, showing that the scaling weight of the defect identity is given by the one-point function of the stress tensor in the presence of the defect. 
Other defect operators were also shown to be in cohomology and their scaling weights were conjectured. Most notably among these one finds the \emph{displacement} supermultiplet. This allows to compute correlation functions in the presence of the defect from chiral algebra, provided one can identify the image in chiral algebra of the defect identity.

\subsection{Gravity dual}

In \cite{Bonetti:2016nma} the question of what is the $AdS$ dual description of the chiral algebra sector was  explored. According to the $AdS/CFT$ correspondence \cite{Maldacena:1997re} $\NN=4$ Super-Yang-Mills (SYM) with gauge group $su(N)$ is dual to IIB string theory on $AdS_5\times S^5$,  with the CFT, $\NN=4$ SYM, living on the boundary of $AdS$.\footnote{See \eg, \cite{Aharony:1999ti} for a review.}   As $N\to \infty$ the string theory is weakly coupled, and at strong coupling 't Hooft coupling, $g_{YM}^2 N$, where $g_{YM}$ is the $su(N)$ $\NN=4$ gauge coupling, the $AdS$ curvature is small and one can trust the supergravity approximation of string theory.

The natural question arises of what is the gravity dual of the protected subsector obtained by passing to the cohomology of $\qq_i$.\footnote{Recall that the 't Hooft coupling is not visible in the chiral algebra.} This subsector should be obtained by a suitable version of supersymmetric localization, using the bulk analog of the boundary supercharge $\qq_i$. The proposal of \cite{Bonetti:2016nma} is that the bulk dual of the chiral algebra in the leading large $N$ limit is a Chern-Simons theory on an $AdS_3$ slice of $AdS_5$, with the gauge algebra being a suitable infinite-dimensional supersymmetric higher-spin algebra.
The authors consider the simplest truncation of supergravity for which a convenient off-shell formalism is available, allowing them to do a localization computation. Namely, they consider in $AdS$ an $\NN=4$  (half-maximal supersymmetry) vector multiplet with gauge algebra $g_f$.\footnote{This is not a consistent truncation of the full supergravity equations of motion, but it is a consistent truncation of the $\qq_i$ cohomology.} According to the AdS/CFT dictionary, the $\NN=2$ SCFT will thus have a flavor symmetry $g_f$, with the corresponding chiral algebra being an AKM current algebra $\hat{g}_f$. The localization computation yields Chern-Simons theory in $AdS_3$ with gauge algebra $g_f$. The case of $\NN=4$ SYM corresponds to taking $g_f=su(2)$, since when viewing an $\NN=4$ SCFT as an $\NN=2$ one, the $su(4)_R$ symmetry appears as a combination of the $su(2)_R$ symmetry of $\NN=2$ SCFTs and an ``extra'' $su(2)_f$ flavor symmetry.

\subsection{Obtaining the chiral algebra from \texorpdfstring{$\Omega$-deformation}{Obtaining the chiral algebra from Omega-deformation}}

Kapustin introduced \cite{Kapustin:2006hi} a topological-holomorphic twist of four-dimensional QFTs placed on $\Sigma \times \CC$, that renders the theory topological on $\Sigma$ and holomorphic on $\CC$. Such a twist is achieved by passing to the cohomology of
\be 
Q_{HT} = \QQ_{-}^1 + \Qt_{2\md}\,,
\ee
under which translations in $\Sigma$ or in the anti-holomorphic direction of $\CC$ are $Q_{HT}-$exact. The algebra of local operators in the $Q_{HT}$ cohomology has a commutative vertex algebra structure, and comes equipped with a Poisson bracket. The same Schur operators that contribute to the chiral algebra described in these lectures are in this cohomology, but now the OPEs are non-singular. In \cite{Oh:2019bgz,Jeong:2019pzg} it was shown how to obtain the cohomology of $\qq_i$ from an $\Omega-$ deformation of this construction.
Taking the SCFT on $\Sigma \times \CC = \mathbb{R}^2 \times \mathbb{C}$, the  $\Omega-$deformation replaces $Q_{HT}$ with a linear combination of Poincar\'{e} and conformal supercharges
\be
Q^{\zeta} =\QQ_-^1 + \Qt_{2\md} + \zeta \left(\St^{2 \md} - \SS_1^-\right)\,,
\ee
which corresponds to a quantization of the Poisson vertex algebra.
For a unitary SCFT the $Q^{\zeta} -$cohomology of local operators is isomorphic to the cohomology of $\qq_{i=1,2}$, thus recovering the chiral algebra described in these lectures \cite{Oh:2019bgz,Jeong:2019pzg}.

%% file: sections/acknowledgments.tex

\acknowledgments

I am greatly indebted to all the collaborators with whom I've worked on chiral algebras of $\NN=2$ SCFTs and from whom I've learned a lot. It is a pleasure to thank Chris Beem, Lorenzo Bianchi, Martina Cornagliotto, Pedro Liendo,  Carlo Meneghelli, Vladimir Mitev, Wolfger Peelaers, Balt van Rees, Volker Schomerus, and most especially Leonardo Rastelli for introducing me and guiding me through the subject. I would also like to thank the organizers of YRIS2020, the other lecturers at the school, and all the students, for making the school very enjoyable, and for many discussions from which these lectures benefited greatly.
I also acknowledge FCT-Portugal under grant PTDC/MAT-OUT/28784/2017.

%% file: sections/A_superalgebra.tex

\section{Conventions and \texorpdfstring{$\NN=2$}{N=2} superconformal algebra}
\label{app:conventions}

\noindent
We raise and lower $su(2)$ indices with epsilon tensor according to $\phi^a = \epsilon^{a b} \phi_b$, $\phi_a = \epsilon_{ab} \phi^b$, and we take $\epsilon_{12}=1$, $\epsilon^{12}=-1$. Going from vector to spinor indices we use the sigma matrices
\be
 \sigma^{\mu}_{\aa \bbd} =(\sigma^a,\ii\mathbb{1})\,, \qquad  (\bar \sigma^{\mu})^{ \aad \bb} =(\sigma^a,-\ii \mathbb{1})\,,
\ee
where $\sigma^a$ are the Pauli matrices. Note that 
\be
- \bar{\sigma}_{\lambda}^{\aad \aa} = \epsilon^{\aa \bb} \epsilon^{\aad \bbd} (\sigma_{\lambda})_{\bb \bbd}\,,
\ee
and so we have to be careful that, except $K^{\aad \aa}$, all fields change from vector to spinor as $\OO_{\aa \aad} = \sigma^\mu_{\aa \aad} P_\mu$. Note also $\OO_\mu = \tfrac12 \bar{\sigma}_\mu^{\aad \aa} \OO_{\aa \aad}$. Round brackets around indices mean symmetrizations with unit strength.

We label operators by their eigenvalues of the Cartans of the symmetry algebra, and so for $\NN=2$ SCFTs they will be labeled by their dimension $\Delta$, Lorentz spin $(j,\jb)$, and R-symmetry representations $su(2)_R\oplus u(1)_r$, and we denote representations by their Dynkin labels.

\subsection{Superconformal algebra}
\label{app:confalgebra}

\subsection*{Conformal algebra}

The conformal transformations and their actions on primary fields read
\be
\begin{alignedat}{3}
&x_\mu \to x_\mu + a_\mu & \phi \to e^{ P_\mu a_\mu } \phi e^{-P_\mu a_\mu} \,, \qquad &\left[ P_\mu, \phi \right] =\partial_\mu \phi\,, \\
&x \to e^\delta x\,, \qquad &\phi \to e^{ \delta \HH} \phi e^{- \delta \HH}\,, \qquad &\left[\HH,\phi\right] =(\Delta_\phi + x \cdot \partial) \phi\,, \\
&x_\mu \to \frac{x_\mu - b_\mu x^2}{1-2 b\cdot x + b^2 x^2}\,, \qquad &\phi \to e^{ b_\mu K_\mu }\phi e^{ -b_\mu K_\mu }\,, \qquad &\left[K_\mu, \phi \right] = -(x^2 \partial_\mu - 2 x_\mu x \cdot \partial - 2 \Delta_\phi x_\mu )\,,\\
&x_\mu \to m_\mu^{\phantom{\mu}\nu} x_\nu\,, &\phi \to e^{ M_{\mu \nu} m^{\mu\nu} } \phi e^{- M_{\mu \nu} m^{\mu\nu}}\,,  \quad & \left[M_{\mu \nu}, \phi \right] = x_\nu \partial_\mu - x_\mu \partial_\nu\,,
\end{alignedat}
\ee
where we took $\phi$ to be a scalar field for simplicity. We can also compute the commutation relations, where we omit those involving $M_{\mu \nu}$ since all commutation relations will be written again in spinor components below,
\be 
\left[ \HH, P_\mu \right] =  P_\mu \,, \qquad \left[\HH,K_\mu \right]=- K_\mu \,, \qquad \left[K_\mu , P_\nu\right] = 2 ( \delta_{\mu \nu} \HH -M_{\mu \nu} )\,.
\ee
We define also the stress tensor as the current generating translations by
\be 
P_\nu (\Sigma)  =: - \int_\Sigma d\Omega_\mu T^{\mu \nu}(x)\,,
\label{eq:Tdef}
\ee
where $d\Omega_\mu = d\Omega \frac{x_\mu}{|x|}$ . Then the Ward identity satisfied by the stress tensor is \eg, \cite{Osborn:1993cr, Simmons-Duffin:2016gjk},
\be 
\partial_\mu T^{\mu \nu}(x) \OO(y)= - \delta^d(x-y) \partial_\mu \OO(x)\,, \qquad T^{\mu \mu}(x) \OO(y) = - \Delta_\OO O(x) \delta^d(x-y)\,,
\ee
which can be checked to be consitent with the commutation relation for $P$ defined above, and is also consitent for dilatations where we have  following \cite{Poland:2010wg}
\be 
\HH = - \int_\Sigma d\Omega_\mu x_\nu T^{\mu \nu} \,.
\label{eq:dilatations}
\ee
This can be checked by integrating the the Ward identities above in a ball surrounding an operator.
These generators have the following properties under conjugation in radial quantization  $\HH^{\dagger} = \HH$, $P_\mu ^{\dagger}  = K_\mu$, $(M_{\mu \nu})^\dagger = -M_{\mu \nu}$, where here and throughout the lectures $\dagger$ means the conjugate in radial quantization.

\subsection*{\texorpdfstring{$\NN=2$}{N=2} superconformal algebra}
\label{app:SCA}

We have modified the $\NN=2$ superconformal algebra of \cite{Beem:2013sza} such that the bosonic generators are as defined above.\footnote{In particular $K_{here}=-2 \ii K^{there}$ and $P_{here}=2 \ii P_{there}$, $\QQ_{here}= \ii \QQ_{there}$, $\SS_{here}=- \ii \SS_{there}$, but  $\Qt_{here}= \Qt_{there}$, $\St_{here}=\St_{there}$.}
Using $P_{\aa \aad} = \sigma^\mu_{\aa \aad} P_\mu$, $K^{ \aad \aa} = \bar{\sigma}_\mu^{\aad \aa} K_\mu$  and
\be 
\MM_{\bb}^{\phantom{\bb} \aa}=-\tfrac14 \bar{\sigma}^{\mu \aad \a} \sigma_{\nu \bb \aad} M_{\mu \nu}\,, \qquad
\Md^{\aad}_{\phantom{\aad} \bbd}= \tfrac14  \bar{\sigma}^{\mu \aad \a} \sigma_{\nu \aa \bbd} M_{\mu \nu}\,, 
\ee
we get the bosonic algebra:\footnote{ Note that this means $(P_{\aa \aad})^\dagger =  K^{\aad \aa}$,  $(\MM_{\bb}^{\phantom{\bb} \aa})^\dagger =  \MM_{\aa}^{\phantom{\aa} \bb}$ and 
 $(\Md^{\aad}_{\phantom{\aad} \bbd})^\dagger =  \Md^{\bbd}_{\phantom{\bbd} \aad}$ . }

\be
\begin{alignedat}{4}
&[\MM_{\aa}^{~\bb},\MM_{\gg}^{\phantom{\gg}\delta}]	&~=~&	\delta_{\gg}^{~\bb}\MM_{\aa}^{~\delta}-\delta_{\aa}^{~\delta}\MM_{\gg}^{~\bb}~,\\
&[\Md^{\aad}_{~\bbd},\Md^{\ggd}_{~\ddd}]			&~=~&	-\delta^{\aad}_{~\ddd}\Md^{\ggd}_{~\bbd}+\delta^{\ggd}_{~\bbd}\Md^{\aad}_{~\ddd}~,\\
&[\MM_{\aa}^{~\bb},\PP_{\gg\ggd}]				&~=~&	\delta_{\gg}^{~\bb}\PP_{\aa\ggd}-\tfrac12\delta_{\aa}^{\phantom{\aa}\bb}\PP_{\gg\ggd}~,\\
&[\Md^{\aad}_{~\bbd},\PP_{\gg\ggd}]				&~=~&	-\delta^{\aad}_{~\ggd}\PP_{\gg\bbd}+\tfrac12\delta^{\aad}_{\phantom{\aad}\bbd}\PP_{\gg\ggd}~,\\
&[\MM_{\aa}^{~\bb},\KK^{\ggd\gg}]				&~=~&	-\delta_{\aa}^{~\gg}\KK^{\ggd\bb}+\tfrac12\delta_{\aa}^{\phantom{\aa}\bb}\KK^{\ggd\gg}~,\\
&[\Md^{\aad}_{~\bbd},\KK^{\ggd\gg}]				&~=~&	\delta^{\ggd}_{~\bbd}\KK^{\aad\gg}-\tfrac12\delta^{\aad}_{\phantom{\aad}\bbd}\KK^{\ggd\gg}~,\\
&[\HH,\PP_{\aa\aad}]							&~=~&	\PP_{\aa\aad}~,\\
&[\HH,\KK^{\aad\aa}]							&~=~&	- \KK^{\aad\aa}~,\\
&[\KK^{\aad\aa},\PP_{\bb\bbd}]					&~=~&	4 \delta_{\bb}^{\phantom{\bb}\aa}\delta^{\aad}_{\phantom{\aad}\bbd}\HH-4\delta_{\bb}^{\phantom{\bb}\aa}\Md^{\aad}_{\phantom{\aad}\bbd}+4\delta^{\aad}_{\phantom{\aad}\bbd}\MM_{\bb}^{\phantom{\bb}\aa}~.
\end{alignedat}
\ee
We take the standard $su(2)_R$ algebra
\be
[\RR^+,\RR^-]=2\RR~,\qquad [\RR,\RR^{\pm}]=\pm\RR^{\pm}\,,
\ee
and we introduce the basis $\RR^\II_{\phantom{1}\JJ}$, with
\be\label{eq:su2u1}
\RR^1_{\phantom{1}2}=\RR^+~,\qquad\RR^2_{\phantom{2}1}=\RR^-~,\qquad\RR^1_{\phantom{1}1}=-\frac14 r+\RR~,\qquad\RR^2_{\phantom{1}2}=-\frac14 r-\RR~,
\ee
where we follow the conventions of \cite{Cordova:2016emh} for the $u(1)_r$ charge,\footnote{This means $r_{\mathrm{here}} =- 2 r_{\text{\cite{Dolan:2002zh,Beem:2013sza}}}$.  Throughout these lectures we use Dynkin labels for denoting representations, in particular a spin $\frac{R}{2}$ representation of $su(2)_R$ will have Dynkin label $R$, and similarly for the spins.} and which obey the commutation relations
\be
[\RR^\II_{\phantom{\II}\JJ},\RR^{k}_{\phantom{k}\LL}]=\delta^k_{\phantom{k}\JJ}\RR^\II_{\phantom{\II}\LL}-\delta^\II_{\phantom{\II}\LL}\RR^k_{\phantom{k}\JJ}~.
\ee

The eight Poincar\'e supercharges $\QQ^\II_{\aa}$, $\Qt_{\II \aad}$ and eight conformal supercharges $(Q_\alpha^\II)^\dagger=S_\II^\alpha$,  $(\tilde{Q}_{\II \aad})^\dagger=\tilde{S}^{\II \aad}$ (where the dagger means conjugation in radial quantization) obey the following algebra
\be\label{eq:4dSCA}
\begin{alignedat}{4}
&\{\QQ_{\aa}^\II,\,\tilde{\QQ}_{\JJ\aad}\}  			&~=~~&\tfrac12	\delta^\II_{\phantom{\II}\JJ} \PP_{\aa\aad}~,\\
&\{\tilde{\SS}^{\II\aad},\,\SS_{\JJ}^{\phantom{\aa}\aa}\} &~=~~&	\tfrac12 \delta^\II_{\phantom{\II}\JJ} \KK^{\aad\aa}~,\\
&\{\QQ_{\aa}^\II,\,\SS^{\phantom{\aa}\bb}_\JJ\}     	&~=~~&	\tfrac12 \delta^\II_{\phantom{\II}\JJ}\delta_{\aa}^{\phantom{\aa}\bb}\HH   + \delta^\II_{\phantom{\II}\JJ} \MM_{\aa}^{\phantom{\aa}\bb}-\delta_\aa^{\phantom{\aa}\bb} \RR^\II_{\phantom{\II}\JJ}~,\\
&\{\tilde{\SS}^{\II\aad},\,\tilde{\QQ}_{\JJ\bbd}\}		&~=~~&	\tfrac12 \delta^\II_{\phantom{\II}\JJ}\delta^{\aad}_{\phantom{\aad}\bbd}\HH - \delta^\II_{\phantom{\II}\JJ} \Md^{\aad}_{\phantom{\aad}\bbd}+\delta^{\aad}_{\phantom{\aad}\bbd} \RR^\II_{\phantom{\II}\JJ}~,\\
&[\MM_{\aa}^{~\bb},\QQ_{\gg}^\II]	&~=~&	\delta_{\gg}^{~\bb} \QQ_{\aa}^\II -\tfrac12\delta_{\aa}^{\phantom{\aa}\bb} \QQ_{\gg}^\II~,\\
&[\Md^{\aad}_{~\bbd},\tilde{\QQ}_{\II \ddd}]			&~=~& -\delta^{\aad}_{~\ddd}\tilde{\QQ}_{\II \bbd} +\tfrac12\delta^{\aad}_{\phantom{\aad}\bbd}\tilde{\QQ}_{\II \ddd}~,\\
&[\MM_{\aa}^{~\bb},\SS_{\II}^{\phantom{\aa}\gg}]				&~=~&	-\delta_{\aa}^{~\gg}\SS_{\II}^{\phantom{\aa}\bb}+\tfrac12\delta_{\aa}^{\phantom{\aa}\bb} \SS_{\II}^{\phantom{\aa}\gg}~,\\
&[\Md^{\aad}_{~\bbd},\tilde{\SS}^{\II\ggd}]				&~=~&	\delta^{\ggd}_{~\bbd}\tilde{\SS}^{\II\aad}-\tfrac12\delta^{\aad}_{\phantom{\aad}\bbd}\tilde{\SS}^{\II\ggd}~,\\
&[\HH,\QQ_{\aa}^\II]							&~=~& 	\tfrac12 \QQ_{\aa}^\II~,\\
&[\HH,\tilde{\QQ}_{\II \aad}]							&~=~& 	\tfrac12 \tilde{\QQ}_{\II \aad}~,\\
&[\HH, \SS_{\II}^{\phantom{\aa}\aa}]							&~=~&	-\tfrac12  \SS_{\II}^{\phantom{\aa}\aa}~,\\
&[\HH, \tilde{\SS}^{\II\aad} ]							&~=~&	-\tfrac12 \tilde{\SS}^{\II\aad} ~,\\
&[\RR^\II_{\phantom{\II}\JJ},\QQ_{\aa}^k]	&~=~&	\delta_{\JJ}^{~k} \QQ_{\aa}^\II -\frac{1}{4} \delta_{\JJ}^{\II} \QQ_{\aa}^k~,\\
&[\RR^\II_{\phantom{\II}\JJ},\tilde{\QQ}_{k \aad}]	&~=~&	-\delta_{k}^{~\II} \tilde{\QQ}_{\JJ \aad} +\frac{1}{4} \delta_{\JJ}^{\II} \tilde{\QQ}_{k \aad}~,\\
&[\KK^{\aad\aa},\QQ_{\bb}^\II]					&~=~&	2 \delta_{\bb}^{\phantom{\bb}\aa}\tilde{\SS}^{\II\aad}~,\\
&[\KK^{\aad\aa},\tilde{\QQ}_{\II \bbd}]					&~=~&	2 \delta_{\bbd}^{\phantom{\bbd}\aad} \SS_{\II}^{\phantom{\aa}\aa}~,\\
&[\PP_{\aa\aad},\SS_{\II}^{\phantom{\aa}\bb}]					&~=~&	- 2 \delta_{\aa}^{\phantom{\aa}\bb}\tilde{\QQ}_{\II \aad}~,\\
&[\PP_{\aa\aad},\tilde{\SS}^{\II\bbd} ]					&~=~&	-2 \delta_{\aad}^{\phantom{\aad}\bbd} \QQ_{\aa}^\II~,
\end{alignedat}
\ee
where unlisted commutation relations vanish.

\subsection*{Charges and currents}

Similarly to the definition of $P_\mu$ and $\HH$ in eqs.~\eqref{eq:Tdef} and \eqref{eq:dilatations} we define currents for the supercharges and R-symmetry charges as
\be
\label{eq:charges}
\begin{split}
&\RR^\II_{\phantom{\II}\JJ} =: - \ii \int\limits_{\Sigma} d\Omega_\mu   (J_{\RR \mu})^{\II}_{\phantom{\II} \JJ}\,, \qquad j^{u(1)_r}_\mu\colonequals  \frac12 (J_{\RR \mu})^{\II}_{\phantom{\II} \II}\,, \qquad
{j_{\mu \JJ}}^\II\colonequals  (J_{\RR \mu})^{\II}_{\phantom{\II} \JJ}- \frac12 \delta^\II_\JJ J_{\RR \mu} \,, \\
&\QQ^{\II}_\aa =: - \int\limits_{\Sigma} d\Omega_\mu  J^{\mu \II}_{\aa} \,, \qquad
\tilde{\QQ}_{\II \aad} =: - \int\limits_{\Sigma} d\Omega_\mu  \bar{J}_{\mu \II \aad} \,.
\end{split}
\ee
The $u(2)_R$ currents defined in this way are such that $j^{u(1)_r}_\mu$ is the current for the $r$ charge and $j_\mu$ for the $su(2)_R$, and the Ward identity for the conservation of the current is
\be 
\partial_\mu  (J_{\RR \mu})^{\II}_{\phantom{\II} \JJ} (x) \OO(y)^k =  \ii \delta(x-y) \left( \delta^k_\JJ \OO^\II(y) - \frac{1}{4}\delta^\II_\JJ \OO^k(y)\right)\,.
\label{eq:currentWard}
\ee

For reference the canonically normalized stress tensor has a two-point function given by (see \eg, \cite{Osborn:1993cr})
\be
\langle T_{\mu \nu}(x) T_{\rho \sigma}(0) \rangle = \frac{40c}{\pi^4 x^8} \mathcal{I}_{\mu \nu, \rho \sigma}(x)\,,\\
\label{eq:STtwopoint}
\ee
where 
\be 
\mathcal{I}_{\mu \nu, \rho \sigma}(x) = \frac12\left(I_{\mu \rho}(x) I_{\nu \rho}(x) + I_{\mu \rho}(x) I_{\nu \sigma}(x)\right) - \frac14 \delta_{\mu \nu}\delta_{\rho\sigma}\,, \qquad I_{\mu \nu}(x) = \delta_{\mu \nu} - 2 \frac{x_\mu x_\nu}{x^2}\,,
\ee
and where $c_{4d}$ is the usual central charge normalized such that a free $\NN=2$ hypermultiplet has $c_{4d}=\tfrac{1}{12}$. 

For the flavor currents we take the charge to be defined as 
\be 
T^A 	\colonequals - \ii \int d\Omega_\mu J_\mu^A \,,
\ee
where $A$ is an adjoint index, and which implies the following Ward identity for a scalar field $\phi_i$
\be 
\partial_\mu J_\mu^A(x) \phi_i (y) \sim \ii (T^A)_i^j \phi_j \delta(x-y)\,,
\ee
and OPE
\be 
J_\mu^A (x) \phi_i(0) \sim \frac{\ii}{2 \pi^2} \frac{(T^A)_i^j \phi_j(0) x_\mu}{x^2}\,.
\ee
The three-point function of flavor currents and its relation to the two-point function are given in eqs.~(2.23), (3.7), (3.9)  and (6.12) of \cite{Osborn:1993cr}.

%% file: Neq2SCFT_VOA.bbl
\providecommand{\href}[2]{#2}\begingroup\raggedright\begin{thebibliography}{10}

\bibitem{Beem:2013sza}
C.~Beem, M.~Lemos, P.~Liendo, W.~Peelaers, L.~Rastelli, and B.~C. van Rees,
  {\it {Infinite Chiral Symmetry in Four Dimensions}},  {\em Commun. Math.
  Phys.} {\bf 336} (2015), no.~3 1359--1433,
  [\href{http://arxiv.org/abs/1312.5344}{{\tt arXiv:1312.5344}}].

\bibitem{Lorenz}
L.~Eberhardt, {\it {Superconformal symmetry and representations}},  {\em to
  appear}.

\bibitem{Mario}
M.~Martone, {\it {The constraining power of Coulomb Branch Geometry}},  {\em to
  appear}.

\bibitem{Abhijit}
A.~Gadde, {\it {Lectures on Superconformal Index}},  {\em to appear}.

\bibitem{Bruno}
B.~Le~Floch, {\it {A slow review of the AGT correspondence}},  {\em to appear}.

\bibitem{Garcia-Etxebarria:2015wns}
I.~García-Etxebarria and D.~Regalado, {\it {$ \mathcal{N}=3 $ four dimensional
  field theories}},  {\em JHEP} {\bf 03} (2016) 083,
  [\href{http://arxiv.org/abs/1512.06434}{{\tt arXiv:1512.06434}}].

\bibitem{Cordova:2016emh}
C.~Cordova, T.~T. Dumitrescu, and K.~Intriligator, {\it {Multiplets of
  Superconformal Symmetry in Diverse Dimensions}},  {\em JHEP} {\bf 03} (2019)
  163, [\href{http://arxiv.org/abs/1612.00809}{{\tt arXiv:1612.00809}}].

\bibitem{Dolan:2002zh}
F.~A. Dolan and H.~Osborn, {\it {On short and semi-short representations for
  four-dimensional superconformal symmetry}},  {\em Annals Phys.} {\bf 307}
  (2003) 41--89, [\href{http://arxiv.org/abs/hep-th/0209056}{{\tt
  hep-th/0209056}}].

\bibitem{Aharony:2013hda}
O.~Aharony, N.~Seiberg, and Y.~Tachikawa, {\it {Reading between the lines of
  four-dimensional gauge theories}},  {\em JHEP} {\bf 08} (2013) 115,
  [\href{http://arxiv.org/abs/1305.0318}{{\tt arXiv:1305.0318}}].

\bibitem{Simmons-Duffin:2016gjk}
D.~Simmons-Duffin, {\it {The Conformal Bootstrap}},  in {\em {Proceedings,
  Theoretical Advanced Study Institute in Elementary Particle Physics: New
  Frontiers in Fields and Strings (TASI 2015): Boulder, CO, USA, June 1-26,
  2015}}, pp.~1--74, 2017.
\newblock \href{http://arxiv.org/abs/1602.07982}{{\tt arXiv:1602.07982}}.

\bibitem{Ferrara:1973yt}
S.~Ferrara, A.~Grillo, and R.~Gatto, {\it {Tensor representations of conformal
  algebra and conformally covariant operator product expansion}},  {\em Annals
  Phys.} {\bf 76} (1973) 161--188.

\bibitem{Polyakov:1974gs}
A.~Polyakov, {\it {Nonhamiltonian approach to conformal quantum field theory}},
   {\em Zh.Eksp.Teor.Fiz.} {\bf 66} (1974) 23--42.

\bibitem{Mack:1975jr}
G.~Mack, {\it {Duality in quantum field theory}},  {\em Nucl. Phys.} {\bf B118}
  (1977) 445--457.

\bibitem{DiFrancesco:1997nk}
P.~Di~Francesco, P.~Mathieu, and D.~Senechal, {\em {Conformal Field Theory}}.
\newblock Graduate Texts in Contemporary Physics. Springer-Verlag, New York,
  1997.

\bibitem{Rattazzi:2008pe}
R.~Rattazzi, V.~S. Rychkov, E.~Tonni, and A.~Vichi, {\it {Bounding scalar
  operator dimensions in 4D CFT}},  {\em JHEP} {\bf 12} (2008) 031,
  [\href{http://arxiv.org/abs/0807.0004}{{\tt arXiv:0807.0004}}].

\bibitem{Poland:2018epd}
D.~Poland, S.~Rychkov, and A.~Vichi, {\it {The Conformal Bootstrap: Theory,
  Numerical Techniques, and Applications}},  {\em Rev. Mod. Phys.} {\bf 91}
  (2019) 015002, [\href{http://arxiv.org/abs/1805.04405}{{\tt
  arXiv:1805.04405}}].

\bibitem{Ribault:2014hia}
S.~Ribault, {\it {Conformal field theory on the plane}},
  \href{http://arxiv.org/abs/1406.4290}{{\tt arXiv:1406.4290}}.

\bibitem{Ginsparg:1988ui}
P.~H. Ginsparg, {\it {APPLIED CONFORMAL FIELD THEORY}},  in {\em {Les Houches
  Summer School in Theoretical Physics: Fields, Strings, Critical Phenomena Les
  Houches, France, June 28-August 5, 1988}}, pp.~1--168, 1988.
\newblock \href{http://arxiv.org/abs/hep-th/9108028}{{\tt hep-th/9108028}}.

\bibitem{Weinberg:1995mt}
S.~Weinberg, {\em {The Quantum theory of fields. Vol. 1: Foundations}}.
\newblock Cambridge University Press, 2005.

\bibitem{Dolan:2001tt}
F.~A. Dolan and H.~Osborn, {\it {Superconformal symmetry, correlation functions
  and the operator product expansion}},  {\em Nucl. Phys.} {\bf B629} (2002)
  3--73, [\href{http://arxiv.org/abs/hep-th/0112251}{{\tt hep-th/0112251}}].

\bibitem{Nirschl:2004pa}
M.~Nirschl and H.~Osborn, {\it {Superconformal Ward identities and their
  solution}},  {\em Nucl. Phys.} {\bf B711} (2005) 409--479,
  [\href{http://arxiv.org/abs/hep-th/0407060}{{\tt hep-th/0407060}}].

\bibitem{Dolan:2004mu}
F.~A. Dolan, L.~Gallot, and E.~Sokatchev, {\it {On four-point functions of
  1/2-BPS operators in general dimensions}},  {\em JHEP} {\bf 09} (2004) 056,
  [\href{http://arxiv.org/abs/hep-th/0405180}{{\tt hep-th/0405180}}].

\bibitem{Shapere:2008zf}
A.~D. Shapere and Y.~Tachikawa, {\it {Central charges of N=2 superconformal
  field theories in four dimensions}},  {\em JHEP} {\bf 09} (2008) 109,
  [\href{http://arxiv.org/abs/0804.1957}{{\tt arXiv:0804.1957}}].

\bibitem{Osborn:1993cr}
H.~Osborn and A.~C. Petkou, {\it {Implications of conformal invariance in field
  theories for general dimensions}},  {\em Annals Phys.} {\bf 231} (1994)
  311--362, [\href{http://arxiv.org/abs/hep-th/9307010}{{\tt hep-th/9307010}}].

\bibitem{Baggio:2012rr}
M.~Baggio, J.~de~Boer, and K.~Papadodimas, {\it {A non-renormalization theorem
  for chiral primary 3-point functions}},  {\em JHEP} {\bf 07} (2012) 137,
  [\href{http://arxiv.org/abs/1203.1036}{{\tt arXiv:1203.1036}}].

\bibitem{Beem:2017ooy}
C.~Beem and L.~Rastelli, {\it {Vertex operator algebras, Higgs branches, and
  modular differential equations}},  {\em JHEP} {\bf 08} (2018) 114,
  [\href{http://arxiv.org/abs/1707.07679}{{\tt arXiv:1707.07679}}].

\bibitem{Bonetti:2018fqz}
F.~Bonetti, C.~Meneghelli, and L.~Rastelli, {\it {VOAs labelled by complex
  reflection groups and 4d SCFTs}},  {\em JHEP} {\bf 05} (2019) 155,
  [\href{http://arxiv.org/abs/1810.03612}{{\tt arXiv:1810.03612}}].

\bibitem{Beem:2019tfp}
C.~Beem, C.~Meneghelli, and L.~Rastelli, {\it {Free Field Realizations from the
  Higgs Branch}},  {\em JHEP} {\bf 09} (2019) 058,
  [\href{http://arxiv.org/abs/1903.07624}{{\tt arXiv:1903.07624}}].

\bibitem{Argyres:2007cn}
P.~C. Argyres and N.~Seiberg, {\it {S-duality in N=2 supersymmetric gauge
  theories}},  {\em JHEP} {\bf 12} (2007) 088,
  [\href{http://arxiv.org/abs/0711.0054}{{\tt arXiv:0711.0054}}].

\bibitem{Nishinaka:2016hbw}
T.~Nishinaka and Y.~Tachikawa, {\it {On 4d rank-one $ \mathcal{N}=3 $
  superconformal field theories}},  {\em JHEP} {\bf 09} (2016) 116,
  [\href{http://arxiv.org/abs/1602.01503}{{\tt arXiv:1602.01503}}].

\bibitem{Lemos:2016xke}
M.~Lemos, P.~Liendo, C.~Meneghelli, and V.~Mitev, {\it {Bootstrapping
  $\mathcal{N}=3$ superconformal theories}},  {\em JHEP} {\bf 04} (2017) 032,
  [\href{http://arxiv.org/abs/1612.01536}{{\tt arXiv:1612.01536}}].

\bibitem{Song:2016yfd}
J.~Song, {\it {Macdonald Index and Chiral Algebra}},  {\em JHEP} {\bf 08}
  (2017) 044, [\href{http://arxiv.org/abs/1612.08956}{{\tt arXiv:1612.08956}}].

\bibitem{Beem:2019snk}
C.~Beem, C.~Meneghelli, W.~Peelaers, and L.~Rastelli, {\it {VOAs and rank-two
  instanton SCFTs}},  \href{http://arxiv.org/abs/1907.08629}{{\tt
  arXiv:1907.08629}}.

\bibitem{Xie:2019zlb}
D.~Xie and W.~Yan, {\it {Schur sector of Argyres-Douglas theory and
  $W$-algebra}},  \href{http://arxiv.org/abs/1904.09094}{{\tt
  arXiv:1904.09094}}.

\bibitem{Buican:2015ina}
M.~Buican and T.~Nishinaka, {\it {On the superconformal index of
  Argyres–Douglas theories}},  {\em J. Phys.} {\bf A49} (2016), no.~1 015401,
  [\href{http://arxiv.org/abs/1505.05884}{{\tt arXiv:1505.05884}}].

\bibitem{DPKR}
L.~Di~Pietro, Z.~Komargodski, and L.~Rastelli, {\it {unpublished notes 2015}},
  .

\bibitem{Ardehali:2015bla}
A.~Arabi~Ardehali, {\it {High-temperature asymptotics of supersymmetric
  partition functions}},  {\em JHEP} {\bf 07} (2016) 025,
  [\href{http://arxiv.org/abs/1512.03376}{{\tt arXiv:1512.03376}}].

\bibitem{DiPietro:2014bca}
L.~Di~Pietro and Z.~Komargodski, {\it {Cardy formulae for SUSY theories in $d
  =$ 4 and $d =$ 6}},  {\em JHEP} {\bf 12} (2014) 031,
  [\href{http://arxiv.org/abs/1407.6061}{{\tt arXiv:1407.6061}}].

\bibitem{Thielemans:1991uw}
K.~Thielemans, {\it {A Mathematica package for computing operator product
  expansions}},  {\em Int. J. Mod. Phys.} {\bf C2} (1991) 787--798.

\bibitem{Krivonos:1995bk}
S.~Krivonos and K.~Thielemans, {\it {A Mathematica package for computing N=2
  superfield operator product expansions}},  {\em Class. Quant. Grav.} {\bf 13}
  (1996) 2899--2910, [\href{http://arxiv.org/abs/hep-th/9512029}{{\tt
  hep-th/9512029}}].

\bibitem{Lemos:2014lua}
M.~Lemos and W.~Peelaers, {\it {Chiral Algebras for Trinion Theories}},  {\em
  JHEP} {\bf 02} (2015) 113, [\href{http://arxiv.org/abs/1411.3252}{{\tt
  arXiv:1411.3252}}].

\bibitem{Choi:2017nur}
J.~Choi and T.~Nishinaka, {\it {On the chiral algebra of Argyres-Douglas
  theories and S-duality}},  {\em JHEP} {\bf 04} (2018) 004,
  [\href{http://arxiv.org/abs/1711.07941}{{\tt arXiv:1711.07941}}].

\bibitem{rastelli_harvard}
C.~Beem and L.~Rastelli, ``{Infinite Chiral Symmetry in Four and Six
  Dimensions}.'' {Seminar at Harvard University by L. Rastelli, November,
  2014.}

\bibitem{Cordova:2015nma}
C.~Cordova and S.-H. Shao, {\it {Schur Indices, BPS Particles, and
  Argyres-Douglas Theories}},  {\em JHEP} {\bf 01} (2016) 040,
  [\href{http://arxiv.org/abs/1506.00265}{{\tt arXiv:1506.00265}}].

\bibitem{Argyres:1995jj}
P.~C. Argyres and M.~R. Douglas, {\it {New phenomena in SU(3) supersymmetric
  gauge theory}},  {\em Nucl. Phys.} {\bf B448} (1995) 93--126,
  [\href{http://arxiv.org/abs/hep-th/9505062}{{\tt hep-th/9505062}}].

\bibitem{Argyres:1995xn}
P.~C. Argyres, M.~R. Plesser, N.~Seiberg, and E.~Witten, {\it {New N=2
  superconformal field theories in four-dimensions}},  {\em Nucl. Phys.} {\bf
  B461} (1996) 71--84, [\href{http://arxiv.org/abs/hep-th/9511154}{{\tt
  hep-th/9511154}}].

\bibitem{Maruyoshi:2016tqk}
K.~Maruyoshi and J.~Song, {\it {Enhancement of Supersymmetry via
  Renormalization Group Flow and the Superconformal Index}},  {\em Phys. Rev.
  Lett.} {\bf 118} (2017), no.~15 151602,
  [\href{http://arxiv.org/abs/1606.05632}{{\tt arXiv:1606.05632}}].

\bibitem{Maruyoshi:2016aim}
K.~Maruyoshi and J.~Song, {\it {$ \mathcal{N}=1 $ deformations and RG flows of
  $ \mathcal{N}=2 $ SCFTs}},  {\em JHEP} {\bf 02} (2017) 075,
  [\href{http://arxiv.org/abs/1607.04281}{{\tt arXiv:1607.04281}}].

\bibitem{Agarwal:2016pjo}
P.~Agarwal, K.~Maruyoshi, and J.~Song, {\it {$ \mathcal{N} $ =1 Deformations
  and RG flows of $ \mathcal{N} $ =2 SCFTs, part II: non-principal
  deformations}},  {\em JHEP} {\bf 12} (2016) 103,
  [\href{http://arxiv.org/abs/1610.05311}{{\tt arXiv:1610.05311}}]. [Addendum:
  JHEP04,113(2017)].

\bibitem{Agarwal:2017roi}
P.~Agarwal, A.~Sciarappa, and J.~Song, {\it {$ \mathcal{N} $ =1 Lagrangians for
  generalized Argyres-Douglas theories}},  {\em JHEP} {\bf 10} (2017) 211,
  [\href{http://arxiv.org/abs/1707.04751}{{\tt arXiv:1707.04751}}].

\bibitem{Benvenuti:2017bpg}
S.~Benvenuti and S.~Giacomelli, {\it {Lagrangians for generalized
  Argyres-Douglas theories}},  {\em JHEP} {\bf 10} (2017) 106,
  [\href{http://arxiv.org/abs/1707.05113}{{\tt arXiv:1707.05113}}].

\bibitem{Beem:2014rza}
C.~Beem, W.~Peelaers, L.~Rastelli, and B.~C. van Rees, {\it {Chiral algebras of
  class S}},  {\em JHEP} {\bf 05} (2015) 020,
  [\href{http://arxiv.org/abs/1408.6522}{{\tt arXiv:1408.6522}}].

\bibitem{Feigin:1990pn}
B.~Feigin and E.~Frenkel, {\it {Quantization of the Drinfeld-Sokolov
  reduction}},  {\em Phys. Lett.} {\bf B246} (1990) 75--81.

\bibitem{deBoer:1993iz}
J.~de~Boer and T.~Tjin, {\it {The Relation between quantum W algebras and Lie
  algebras}},  {\em Commun. Math. Phys.} {\bf 160} (1994) 317--332,
  [\href{http://arxiv.org/abs/hep-th/9302006}{{\tt hep-th/9302006}}].
  [,317(1993)].

\bibitem{Arakawa:2018egx}
T.~Arakawa, {\it {Chiral algebras of class $\mathcal{S}$ and Moore-Tachikawa
  symplectic varieties}},  \href{http://arxiv.org/abs/1811.01577}{{\tt
  arXiv:1811.01577}}.

\bibitem{Beem:2020pry}
C.~Beem and W.~Peelaers, {\it {Argyres-Douglas Theories in Class S Without
  Irregularity}},  \href{http://arxiv.org/abs/2005.12282}{{\tt
  arXiv:2005.12282}}.

\bibitem{Beem:2013qxa}
C.~Beem, L.~Rastelli, and B.~C. van Rees, {\it {The $\mathcal N=4$
  Superconformal Bootstrap}},  {\em Phys. Rev. Lett.} {\bf 111} (2013) 071601,
  [\href{http://arxiv.org/abs/1304.1803}{{\tt arXiv:1304.1803}}].

\bibitem{Beem:2016wfs}
C.~Beem, L.~Rastelli, and B.~C. van Rees, {\it {More ${\mathcal N}=4$
  superconformal bootstrap}},  {\em Phys. Rev.} {\bf D96} (2017), no.~4 046014,
  [\href{http://arxiv.org/abs/1612.02363}{{\tt arXiv:1612.02363}}].

\bibitem{Liendo:2015ofa}
P.~Liendo, I.~Ramirez, and J.~Seo, {\it {Stress-tensor OPE in $ \mathcal{N}=2 $
  superconformal theories}},  {\em JHEP} {\bf 02} (2016) 019,
  [\href{http://arxiv.org/abs/1509.00033}{{\tt arXiv:1509.00033}}].

\bibitem{Lemos:2015orc}
M.~Lemos and P.~Liendo, {\it {$\mathcal{N}=2$ central charge bounds from $2d$
  chiral algebras}},  {\em JHEP} {\bf 04} (2016) 004,
  [\href{http://arxiv.org/abs/1511.07449}{{\tt arXiv:1511.07449}}].

\bibitem{Cornagliotto:2017dup}
M.~Cornagliotto, M.~Lemos, and V.~Schomerus, {\it {Long Multiplet Bootstrap}},
  {\em JHEP} {\bf 10} (2017) 119, [\href{http://arxiv.org/abs/1702.05101}{{\tt
  arXiv:1702.05101}}].

\bibitem{Beem:2018duj}
C.~Beem, {\it {Flavor Symmetries and Unitarity Bounds in ${\mathcal N}=2$
  Superconformal Field Theories}},  {\em Phys. Rev. Lett.} {\bf 122} (2019),
  no.~24 241603, [\href{http://arxiv.org/abs/1812.06099}{{\tt
  arXiv:1812.06099}}].

\bibitem{Maldacena:2011jn}
J.~Maldacena and A.~Zhiboedov, {\it {Constraining Conformal Field Theories with
  A Higher Spin Symmetry}},  {\em J. Phys.} {\bf A46} (2013) 214011,
  [\href{http://arxiv.org/abs/1112.1016}{{\tt arXiv:1112.1016}}].

\bibitem{Alba:2013yda}
V.~Alba and K.~Diab, {\it {Constraining conformal field theories with a higher
  spin symmetry in d=4}},  \href{http://arxiv.org/abs/1307.8092}{{\tt
  arXiv:1307.8092}}.

\bibitem{ZhuThesis}
Y.~Zhu, {\it {Vertex operator algebras, elliptic functions and modular forms}},
   {\em ProQuest LLC, Ann Arbor, MI, 1990 Thesis (Ph.D.)--Yale University.}

\bibitem{arakawa2010remark}
T.~Arakawa, {\it A remark on the $c_2$-cofiniteness condition on vertex
  algebras},  2010.

\bibitem{Beem:2014kka}
C.~Beem, L.~Rastelli, and B.~C. van Rees, {\it {$ \mathcal{W} $ symmetry in six
  dimensions}},  {\em JHEP} {\bf 05} (2015) 017,
  [\href{http://arxiv.org/abs/1404.1079}{{\tt arXiv:1404.1079}}].

\bibitem{Chester:2014mea}
S.~M. Chester, J.~Lee, S.~S. Pufu, and R.~Yacoby, {\it {Exact Correlators of
  BPS Operators from the 3d Superconformal Bootstrap}},  {\em JHEP} {\bf 03}
  (2015) 130, [\href{http://arxiv.org/abs/1412.0334}{{\tt arXiv:1412.0334}}].

\bibitem{Beem:2016cbd}
C.~Beem, W.~Peelaers, and L.~Rastelli, {\it {Deformation quantization and
  superconformal symmetry in three dimensions}},  {\em Commun. Math. Phys.}
  {\bf 354} (2017), no.~1 345--392,
  [\href{http://arxiv.org/abs/1601.05378}{{\tt arXiv:1601.05378}}].

\bibitem{Dedushenko:2019mzv}
M.~Dedushenko, {\it {From VOAs to short star products in SCFT}},
  \href{http://arxiv.org/abs/1911.05741}{{\tt arXiv:1911.05741}}.

\bibitem{Pan:2019shz}
Y.~Pan and W.~Peelaers, {\it {Deformation quantizations from vertex operator
  algebras}},  \href{http://arxiv.org/abs/1911.09631}{{\tt arXiv:1911.09631}}.

\bibitem{Dedushenko:2019mnd}
M.~Dedushenko and Y.~Wang, {\it {4d/2d $\rightarrow $ 3d/1d: A song of
  protected operator algebras}},  \href{http://arxiv.org/abs/1912.01006}{{\tt
  arXiv:1912.01006}}.

\bibitem{Drukker:2009sf}
N.~Drukker and J.~Plefka, {\it {Superprotected n-point correlation functions of
  local operators in N=4 super Yang-Mills}},  {\em JHEP} {\bf 04} (2009) 052,
  [\href{http://arxiv.org/abs/0901.3653}{{\tt arXiv:0901.3653}}].

\bibitem{Liendo:2016ymz}
P.~Liendo and C.~Meneghelli, {\it {Bootstrap equations for $ \mathcal{N} $ = 4
  SYM with defects}},  {\em JHEP} {\bf 01} (2017) 122,
  [\href{http://arxiv.org/abs/1608.05126}{{\tt arXiv:1608.05126}}].

\bibitem{defectLCW}
C.~Beem, W.~Peelaers, and L.~Rastelli, {\it {unpublished work}}, .

\bibitem{Cordova:2017mhb}
C.~Cordova, D.~Gaiotto, and S.-H. Shao, {\it {Surface Defects and Chiral
  Algebras}},  {\em JHEP} {\bf 05} (2017) 140,
  [\href{http://arxiv.org/abs/1704.01955}{{\tt arXiv:1704.01955}}].

\bibitem{Nishinaka:2018zwq}
T.~Nishinaka, S.~Sasa, and R.-D. Zhu, {\it {On the Correspondence between
  Surface Operators in Argyres-Douglas Theories and Modules of Chiral
  Algebra}},  {\em JHEP} {\bf 03} (2019) 091,
  [\href{http://arxiv.org/abs/1811.11772}{{\tt arXiv:1811.11772}}].

\bibitem{Pan:2017zie}
Y.~Pan and W.~Peelaers, {\it {Chiral Algebras, Localization and Surface
  Defects}},  {\em JHEP} {\bf 02} (2018) 138,
  [\href{http://arxiv.org/abs/1710.04306}{{\tt arXiv:1710.04306}}].

\bibitem{Bianchi:2019sxz}
L.~Bianchi and M.~Lemos, {\it {Superconformal surfaces in four dimensions}},
  \href{http://arxiv.org/abs/1911.05082}{{\tt arXiv:1911.05082}}.

\bibitem{Bonetti:2016nma}
F.~Bonetti and L.~Rastelli, {\it {Supersymmetric localization in AdS$_{5}$ and
  the protected chiral algebra}},  {\em JHEP} {\bf 08} (2018) 098,
  [\href{http://arxiv.org/abs/1612.06514}{{\tt arXiv:1612.06514}}].

\bibitem{Maldacena:1997re}
J.~M. Maldacena, {\it {The Large N limit of superconformal field theories and
  supergravity}},  {\em Int. J. Theor. Phys.} {\bf 38} (1999) 1113--1133,
  [\href{http://arxiv.org/abs/hep-th/9711200}{{\tt hep-th/9711200}}]. [Adv.
  Theor. Math. Phys.2,231(1998)].

\bibitem{Aharony:1999ti}
O.~Aharony, S.~S. Gubser, J.~M. Maldacena, H.~Ooguri, and Y.~Oz, {\it {Large N
  field theories, string theory and gravity}},  {\em Phys. Rept.} {\bf 323}
  (2000) 183--386, [\href{http://arxiv.org/abs/hep-th/9905111}{{\tt
  hep-th/9905111}}].

\bibitem{Kapustin:2006hi}
A.~Kapustin, {\it {Holomorphic reduction of N=2 gauge theories, Wilson-'t Hooft
  operators, and S-duality}},  \href{http://arxiv.org/abs/hep-th/0612119}{{\tt
  hep-th/0612119}}.

\bibitem{Oh:2019bgz}
J.~Oh and J.~Yagi, {\it {Chiral algebras from $\Omega$-deformation}},  {\em
  JHEP} {\bf 08} (2019) 143, [\href{http://arxiv.org/abs/1903.11123}{{\tt
  arXiv:1903.11123}}].

\bibitem{Jeong:2019pzg}
S.~Jeong, {\it {SCFT/VOA correspondence via $\Omega$-deformation}},  {\em JHEP}
  {\bf 10} (2019) 171, [\href{http://arxiv.org/abs/1904.00927}{{\tt
  arXiv:1904.00927}}].

\bibitem{Poland:2010wg}
D.~Poland and D.~Simmons-Duffin, {\it {Bounds on 4D Conformal and
  Superconformal Field Theories}},  {\em JHEP} {\bf 05} (2011) 017,
  [\href{http://arxiv.org/abs/1009.2087}{{\tt arXiv:1009.2087}}].

\end{thebibliography}\endgroup
